\documentclass[11pt,a4paper]{article}
\usepackage{jheppub}

\usepackage{epsfig}

\usepackage{amsmath, amsthm, amsfonts, amssymb, latexsym}

\usepackage[caption=false]{subfig}

\usepackage{url}
\usepackage{ifpdf}

\def\beq{\begin{equation}}
\def\eeq{\end{equation}}
\def\tu{{\bar u}}
\def\tv{{\bar v}}
\def\tx{{\bar x}}

\def\trho{{\bar \rho}}
\def\th{{\bar h}}

\def\bequ{\begin{equation}}
\def\eequ{\end{equation}}

\arxivnumber{1105.2298, CERN-PH-TH/2011-108}

\title{\Large Radiation from a $D$-dimensional collision of shock waves: \\ first order perturbation theory}

\author[a]{Carlos Herdeiro,} \author[a,b]{Marco O. P. Sampaio,} \author[a]{Carmen Rebelo}
\affiliation[a]{Departamento de F\'\i sica da Universidade de Aveiro and I3N \\ 
Campus de Santiago, 3810-183 Aveiro, Portugal} 
\affiliation[b]{CERN, Physics Department, Theory Division, \\ 
CH-1211 Geneva 23, Switzerland} 

\emailAdd{herdeiro@ua.pt}

 \keywords{Black holes, Large extra dimensions}

\abstract{ 
We study the spacetime obtained by superimposing two equal Aichelburg-Sexl shock waves in $D$ dimensions traveling, head-on, in opposite directions. Considering the collision in a boosted frame, one shock becomes stronger than the other, and a perturbative framework to compute the metric in the future of the collision is setup. The geometry is given, in first order perturbation theory, as an integral solution, in terms of initial data on the null surface where the strong shock has support. We then extract the radiation emitted in the collision by  using a D-dimensional generalisation of the Landau-Lifschitz pseudo-tensor and compute the percentage of the initial centre of mass energy $\epsilon$ emitted as gravitational waves. In $D=4$ we find $\epsilon=25.0\%$, in agreement with the result of D'Eath and Payne \cite{D'Eath:1992hb}. As $D$ increases, this percentage increases monotonically, reaching $40.0\%$ in $D=10$. Our result is always within the bound obtained from apparent horizons by Penrose, in $D=4$, yielding $29.3\%$, and Eardley and Giddings \cite{Eardley:2002re}, in $D> 4$, which  also increases monotonically with dimension, reaching $41.2\%$ in $D=10$. We also present the wave forms and provide a physical interpretation for the observed peaks, in terms of the null generators of the shocks.}

\begin{document}
\maketitle




\section{Introduction}

The ongoing physics runs at the Large Hadron Collider (LHC) with 7 TeV centre of mass energy, are starting to set bounds on physics beyond the standard model. For the particular case of TeV gravity models \cite{ArkaniHamed:1998rs}, microscopic black hole formation and evaporation is predicted for scattering experiments with partonic centre of mass energy beyond the TeV \cite{Dimopoulos:2001hw,Giddings:2001bu}. The first set of bounds for these models was recently released by the CMS~\cite{Khachatryan:2010wx} and ATLAS~\cite{ATLAS-CONF-2011-065, ATLAS-CONF-2011-068} collaborations. The current analysis for the $7$~TeV data, is however extremely dependent on regions of parameter space where (at best) black holes with masses close to the unknown Planckian regime would be produced (see \cite{Park:2011je} for a discussion). Otherwise if strong cuts are imposed, such that the objects produced are in the semi-classical regime (which is the calculable one), the cross-sections become negligible at $7$~TeV, and only after the upgrade of the beam energy to 14 TeV, planned to take place in 2013, will the scenario be properly tested. Any improvement in the phenomenology of these models is therefore quite timely.

Two event generators, \textsc{charybdis2} \cite{Frost:2009cf} and \textsc{blackmax} \cite{Dai:2007ki}
are being used at the LHC to look for signatures of black hole production and evaporation.  The event rates they produce depend sensitively on various input parameters, amongst which two fundamental quantities for the modeling of the black hole production phase are: the critical impact parameter  for black hole formation in parton-parton scattering; and the energy lost into gravitational radiation in this process. The latter, if large, could be an important signature for discovery or exclusion, if dominating the distribution of missing energy and if calculated with enough precision in the semi-classical regime. Furthermore, if the experimental searches continue to rely on events with little (or under-estimated) missing energy as in~\cite{Khachatryan:2010wx,Aad:2009wy,ATLAS-CONF-2011-065}, then this information will be crucial to determine if there is enough phase space left with little missing energy. 

As first argued by t'Hooft \cite{'tHooft:1987rb}, classical gravity, described by general relativity, should dominate transplanckian scattering. Following this rationale, the best estimates, so far, for the two mentioned quantities, come from apparent horizon computations, initiated by Penrose in four spacetime dimensions. He found an apparent horizon on the past light cone of the collision between two shock waves, representing the gravitational field of two particles boosted to the speed of light. Thus, Penrose concluded that no more than 29.3\% of the centre of mass energy could be emitted in gravitational radiation in a head-on-collision.

This analysis was generalised to $D$ dimensions by Eardley and Giddings \cite{Eardley:2002re}. They obtained the bound on the percentage of the centre of mass energy, $\epsilon_{\rm radiated}$, emitted in a head on collision, which increases with dimension and is given by
\begin{equation}
\epsilon_{\rm radiated}\le 1-\frac{1}{2}\left(\frac{D-2}{2}\frac{\Omega_{D-2}}{\Omega_{D-3}}\right)^{\frac{1}{D-2}} \ , 
\end{equation}
where $\Omega_n$ is the volume of the unit $n$-sphere. This bound increases monotonically with dimension approaching $50\%$ in the limit of infinite $D$. Using the results in \cite{Eardley:2002re} for non-head on collisions, the critical impact parameters were numerically evaluated by Nambu and Yoshino \cite{Yoshino:2002tx}. According to their estimate, a black hole will form if at least half of the effective gravitational radius of each of the colliding particles overlap. The state of the art results were obtained by Yoshino and Rychkov \cite{Yoshino:2005hi}, who found an apparent horizon on the future light cone of the collision. Their analysis coincides with the one in \cite{Eardley:2002re} for the energy emitted in head on collisions; but the critical impact parameters for black hole formation become larger, yielding a larger cross section for black hole formation. Events generated by \textsc{charybdis2} rely on a selection of random points from the configuration space allowed by the bounds determined in \cite{Yoshino:2005hi}.

The event horizon, however, is located outside the apparent horizon, and the latter will evolve and settle to the former in the future of the collision. Thus, both the critical impact parameter and $\epsilon_{\rm radiated}$ should be exactly determined by knowledge of the metric in the future of the collision~\footnote{Another attempt to extract the radiation using test particle interactions in Minkowski space was presented in~\cite{Gal'tsov:2009zi,Constantinou:2011ju}.}. There are in principle two methods to compute it, both of which have been developed in four spacetime dimensions.

Modeling the gravitational field of the colliding partons as shock waves, one may use  the Aichelburg-Sexl metric \cite{Aichelburg:1970dh} to describe the process. Using a method first developed by D'Eath \cite{D'Eath:1976ri} and D'Eath and Payne \cite{D'Eath:1992hb,D'Eath:1992hd,D'Eath:1992qu}, one may  compute the metric to the future of the collision analytically, as a perturbative expansion. In four dimensions, second (first) order perturbation theory yields 16.3\% (25.0\%) for the energy emitted in gravitational waves in a head on collision.

Modeling the gravitational field of the colliding partons as highly boosted black holes, one may use numerical relativity techniques to perform fully non-linear numerical simulations of the collisions. In four dimensions the high energy collision of two black holes was performed by Sperhake et. al. \cite{Sperhake:2008ga}, obtaining that  14$\pm$3\% of the energy was converted into gravitational radiation. This overlaps with the result obtained by D'Eath and Payne in second order perturbation theory \cite{D'Eath:1992qu}, which indicates that such an approximation provides a good estimate. Moreover, introducing an impact parameter \cite{Sperhake:2009jz}, one observes that the critical impact parameter increases from about 0.8 in \cite{Yoshino:2005hi} to about 1.2, in units of the Schwarzschild radius of the centre of mass energy. This is an increase of about 50\% which, if verified in the higher dimensional case as well, leads to significantly larger cross-sections, thus becoming easier to exclude regions in the model parameter space. 

Phenomenologically interesting TeV gravity models occur in dimension $D\ge 6$. It would therefore be useful to develop each of the two above methods in higher dimensions. Numerical relativity for higher dimensional spacetimes is being developed (see eg.  \cite{Zilhao:2010sr,Yoshino:2009xp}), and the first ever results for black hole collisions in higher dimensions (albeit low energy collisions) have been produced \cite{Witek:2010xi,Witek:2010az}.  In this paper we shall  perform a higher dimensional collision of shock waves and determine, for even $D$, the energy radiated by computing the geometry to the future of the collision to first order in perturbation theory, building up on the technique of  \cite{D'Eath:1992hb}. The results we obtain for the radiated energy  are summarised in the following table, where the apparent horizon bounds obtained in \cite{Eardley:2002re} are also given for comparison:
\begin{center}
\begin{tabular}{||  c || c | c | c | c | c | c | c |}
\hline			
   Spacetime dimension & 4 & 5 & 6 & 7 & 8 & 9 & 10\\
  \hline
  Apparent horizon bound (\%)  & 29.3 & 33.5 & 36.1 & 37.9 & 39.3 & 40.4 & 41.2\\
  \hline
  First order perturbation theory (\%)  & 25.0 & -- & 33.3 & -- & 37.5 & -- & 40.0 \\
\hline  
\end{tabular}
\end{center}
Like the apparent horizon bound, our results indicate that the energy radiated increases monotonically with $D$. Moreover, our result is always below the bound, as expected. If this reduction trend is verified by higher order perturbation theory (or other methods, such as numerical black hole collisions) this result has phenomenological implications. More energy lost into gravitational radiation means a less massive final black hole is produced. Thus, this result suggests that the final black hole will be more massive than previously estimated, making it more consistent with the semi-classical analysis used for estimating the potentially observable Hawking radiation. 

This paper is organised as follows. In Section \ref{section2} we discuss the spacetime geometry of one and two shock waves (before the collision) in both Brinkmann and Rosen coordinates. In Section \ref{sec3} the perturbative computation is setup and the integral solution for the metric in the future of the collision is provided. Some details concerning the derivation of this solution are provided in Appendix \ref{apa}. In Section \ref{section4} the extraction of gravitational radiation is studied. A formula in $D$ dimensions is first derived in terms of the metric perturbation in the future of the collision starting from the Landau-Lifschitz pseudo tensor (in $D$ dimensions). Then, using results from Appendix \ref{apb}, a workable expression is obtained to extract the energy carried out by the gravitational radiation (formulas \eqref{rad1} and \eqref{rad2}). These expressions are evaluated numerically to yield the results presented in the table above. The wave forms are exhibited and an interpretation is given for the domain where they have support and peak. We also discuss the approximations used in our derivation and further comment on our results in Section \ref{section5}.

\section{One D-dimensional shock wave in Rosen form}
\label{section2}
The D-dimensional Tangherlini solution with mass $M$ is \cite{Tangherlini:1963bw}
\begin{equation}
ds^2 = -\left(1- \frac{16\pi G_D M}{ (D-2) \Omega_{D-2}}\frac{1}{r^{D-3}}\right)
dt^2 + \left(1- \frac{16\pi G_D M}{(D-2) \Omega_{D-2}}\frac{1}{r^{D-3}}
\right)^{-1} dr^2 + r^2 d\Omega_{D-2}^2\ ,
\end{equation}
where $d\Omega_{D-2}^2$ and $\Omega_{D-2}$ are the line element and volume
of the unit $D-2$ sphere. The Aichelburg-Sexl
solution \cite{Aichelburg:1970dh} 
is found by boosting this black hole and then taking simultaneously the limit of infinite boost and vanishing mass, keeping the total energy $\mu$ fixed.  We use a coordinate system $(u,v,x^i)$,
where the retarded and advanced times $(u,v)$ are $(t-z,t+z)$ in terms
of Minkowski coordinates, and $x^i$ are the remaining Cartesian coordinates on the plane of the shock,
$i=1\ldots D-2$, such that the transverse radius is ${\rho}=\sqrt{x^ix_i}$. The resulting geometry for a particle moving in the $+z$ direction is
\begin{equation}
ds^2 = -du dv + d\rho^{2} + \rho^2 d {\Omega}^2_{D-3}+\kappa \Phi(\rho) \delta(u) du^2\ ,\label{AiSe}
\end{equation}
where $\kappa\equiv 8\pi G_D \mu/\Omega_{D-3}$.
The function $\Phi$ depends only on $\rho$ and takes the form \cite{Eardley:2002re}
\begin{equation}
\Phi(\rho)=\left\{
\begin{array}{ll}
 -2\ln(\rho)\ , &  D=4\  \vspace{2mm}\\
\displaystyle{ \frac{2}{(D-4)\rho^{D-4}}}\ , & D>4\ \label{phidef}
\end{array} \right. \ .
\end{equation}

The coordinates used in \eqref{AiSe} are of Brinkmann type \cite{Brinkmann}. Presenting the shock wave in this chart, geodesics and their tangent vectors appear discontinuous across the shock.  One can, however, introduce a new coordinate system defined by \cite{Eardley:2002re}
\begin{eqnarray}
u &=& \tu\ ,\nonumber \\
v &=& \tv+\kappa\theta(\tu)\left(\Phi + \frac{\kappa \tu (\bar{\nabla}\Phi)^2}{4}\right)\ = \tv+\kappa\theta(\tu)\left(\Phi + \frac{\kappa \tu \Phi'^2}{4}\right),\nonumber \\
x^i&=& \tx^i + \kappa\frac{\tu}{2} \bar{\nabla}_i \Phi(\tx)\theta(\tu) \Rightarrow \left\{\begin{array}{rcl} \rho&=& \trho\Big(1+ \frac{\kappa \tu \, \theta(\tu)}{2 \bar \rho}\Phi'\Big)  \\
\phi_{a}&=&\bar \phi_a  \end{array}\right.
\ , \label{ct}
\end{eqnarray}
where $\theta$ is the Heaviside step function and $\Phi$ and its derivative $\Phi'$ are evaluated at $\bar\rho$. In this new chart both geodesics and their tangents are continuous across the shock
at $\tu=u=0$;
$\phi_a$ are the angles on the $(D-3)$-sphere and $a=1...D-3$.
Using~\eqref{ct} the metric becomes \cite{D'Eath:1976ri,Rychkov:2004sf,Yoshino:2005hi}
\begin{equation}
ds^2 = -d\tu d\tv + \Big(1+\dfrac{\kappa \tu \theta(\tu)}{2}\Phi''\Big)^2d\trho^{2} + \trho^2\Big(1+ \frac{\kappa \tu \, \theta(\tu)}{2 \bar \rho}\Phi'\Big)^2 d\bar{\Omega}^2_{D-3} \ ,\label{newmetric}
\end{equation}
which we dub the  \textit{Rosen form} \cite{Rosen} of the shock wave. The geometry for an identical shock wave traveling in the  $-z$ direction is obtained by changing $z\rightarrow -z$ in~\eqref{AiSe} or equivalently exchange $\bar{v}\leftrightarrow \bar{u}$ in~\eqref{newmetric}. Following \cite{D'Eath:1992hb} it is simple to see that in a boosted frame (moving with respect to the $({u},{v})$ chart with velocity $\beta$ in the $-z$ direction), the oppositely directed shock waves keep their form, but with new energy parameters, respectively,
 \begin{equation}
\kappa \rightarrow e^\alpha \kappa\equiv \nu \ , \qquad \kappa \rightarrow e^{-\alpha} \kappa\equiv \lambda \ ,
\end{equation}
where $e^\alpha=\sqrt{(1+\beta)/(1-\beta)}$.

\begin{figure}[t]
\begin{center}
\begin{picture}(0,0)
\put(72,90){$t$}
\put(125,73){$z$}
\put(115,61){$\theta=0$}
\put(0,73){$\theta=\pi$}
\put(289,90){$t$}
\put(341,73){$z$}
\put(334,61){$\theta=0$}
\put(219,73){$\theta=\pi$}
\end{picture}
\includegraphics[scale=0.6]{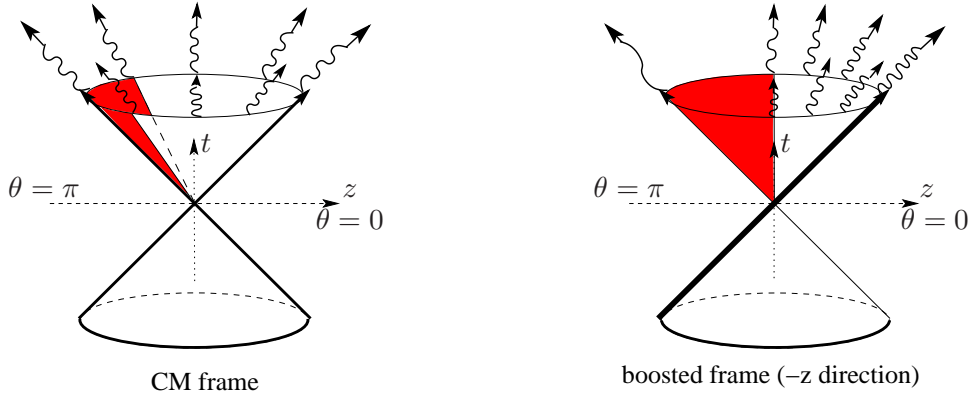}
\end{center}
\caption{Spacetime diagram of the collision in the centre of mass (left panel) and boosted (right panel) frames. The $x^i$ directions are orthogonal to the $t$ and $z$ axis.  In the centre of mass frame, the radiation is symmetric under $z\rightarrow -z$. Only the radiation that propagates along a small solid angle around $\theta=\pi$  in the centre of mass frame (red area, delimited by the polar angle $\tan \theta=\sqrt{\beta^{-2}-1}$) is still propagating in the $-z$ direction in the boosted frame. All the radiation in this solid angle is redshifted in the boosted frame.}
\label{spacetime}
\end{figure}

By causality, the spacetime where two shock waves travel in opposite directions along the $z$ coordinate, is described everywhere by superimposing the two shock wave metrics, except in the future light cone of the collision. Thus, in the boosted frame the spacetime geometry reads
\begin{multline}ds^2 = -d\tu d\tv + \left[\Big(1+\dfrac{\nu \tu \theta(\tu)}{2}\Phi''\Big)^2+ \Big(1+\dfrac{\lambda \tv \theta(\tv)}{2}\Phi''\Big)^2-1\right]d\trho^{2} \\
+ \trho^2\left[\Big(1+ \frac{\nu \tu \, \theta(\tu)}{2 \bar \rho}\Phi'\Big)^2 
 + \Big(1+ \frac{\lambda \tv \, \theta(\tv)}{2 \bar \rho}\Phi'\Big)^2-1\right] d\Omega^2_{D-3}
 \ ,\label{collision}
\end{multline}
which is valid everywhere except in the future light cone of $\tu=\tv=0$. As a consequence of the collision a strong burst of gravitational radiation (followed by a tail) is expected to be emitted. An illustration of this signal in the centre of mass/energy and in the boosted frame is presented in Fig. \ref{spacetime}.

\section{The perturbative metric in the future of the collision}
\label{sec3}
We shall now compute the geometry in the future of the collision to first order in perturbation theory. The main idea is that in the boosted frame one shock wave can carry much more energy than the other, since
\[
\frac{\nu}{\lambda}=e^{2\alpha}=\frac{1+\beta}{1-\beta}\ \ \Rightarrow \ \ \lim_{\beta\rightarrow 1}\frac{\nu}{\lambda}= \infty \ . \] 
Thus, one may face the wave traveling in the $-z$ direction as a small perturbation of the wave traveling in the $+z$ direction. Since the geometry of the latter is flat for $\tu>0$, we make a perturbative expansion of the Einstein equations around flat space,  in order to compute the metric in the future of the collision ($\tv,\tu>0$). To first order this amounts to solving the linearised Einstein equations around flat Minkowski spacetime, subject to a boundary condition given by the metric \eqref{collision} in the limit $u=0^+$.

\begin{figure}[t]
\begin{center}
\includegraphics[scale=0.4]{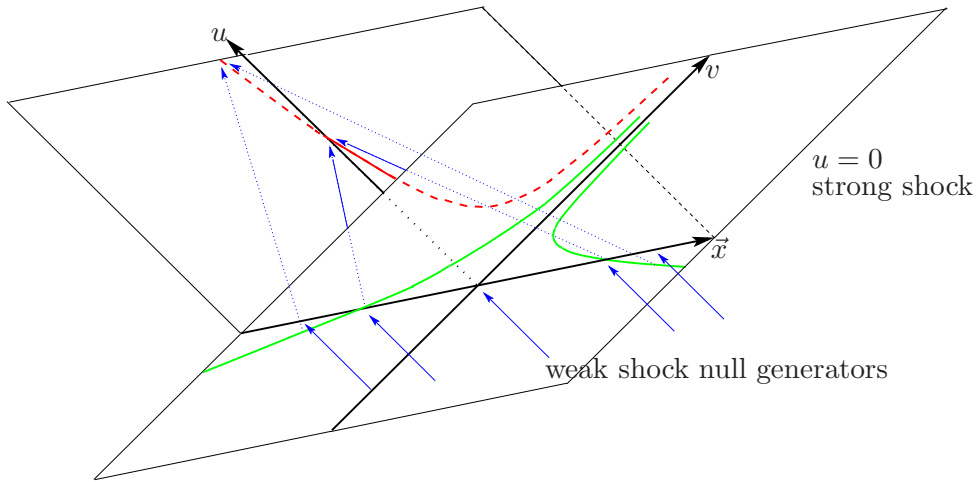}
\begin{picture}(0,0)
\put(-280,166){${u}$}
\put(-95,152){${v}$}
\put(-93,84){${\vec{x}}$}
\put(-155,39){${\rm weak \ shock \ null \ generators}$}
\put(-55,118){$u=0$}
\put(-55,108){${\rm strong \ shock}$}
\end{picture}
\end{center}
\caption{Evolution of the weak shock null generators (blue arrows) from the viewpoint of Brinkmann coordinates in the boosted frame in $D=4$. For $u<0$ they are at $v=0$; then the generators undergo a discontinuity in $v$ at $u=0$, which is $\rho$ dependent and negative for $\rho>1$. They jump to the collision surface (green lines, at $u=0$ and $v=-\sqrt{2}\ln \rho$), gain shear and focus along the caustic (red line).}
\label{focusing4}
\end{figure}

\begin{figure}[th]
\begin{center}
\includegraphics[scale=0.4]{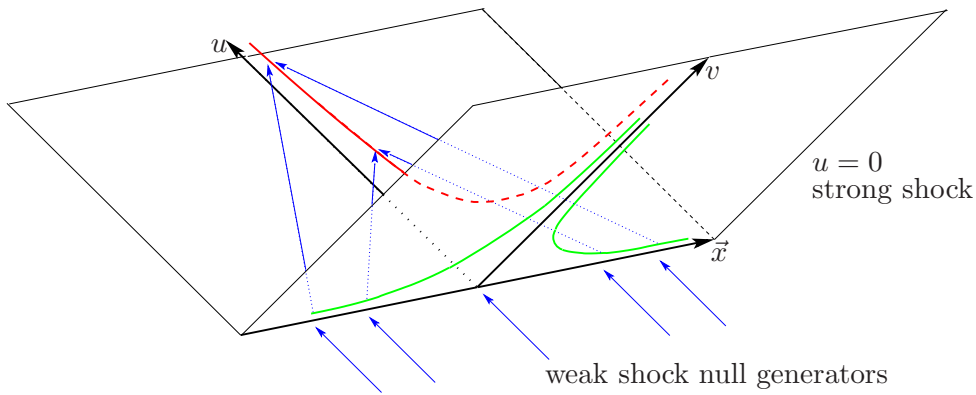}
\begin{picture}(0,0)
\put(-280,129){${u}$}
\put(-95,119){${v}$}
\put(-93,50){${\vec{x}}$}
\put(-155,4){${\rm weak \ shock \ null \ generators}$}
\put(-55,84){$u=0$}
\put(-55,74){${\rm strong \ shock}$}
\end{picture}
\end{center}
\caption{Evolution of the weak shock null generators (blue arrows) from the viewpoint of Brinkmann coordinates in the boosted frame in $D>4$. For $u<0$ they are at $v=0$; then the generators undergo a discontinuity in $v$ at $u=0$, which is $\rho$ dependent but always positive. They jump to the collision surface (green lines, at $u=0, v=\sqrt{2}/[(D-4)\rho^{D-4}]$), gain shear and focus along the caustic (red line).}
\label{focusing}
\end{figure}

\begin{figure}[t]
\begin{center}
\includegraphics[scale=0.7]{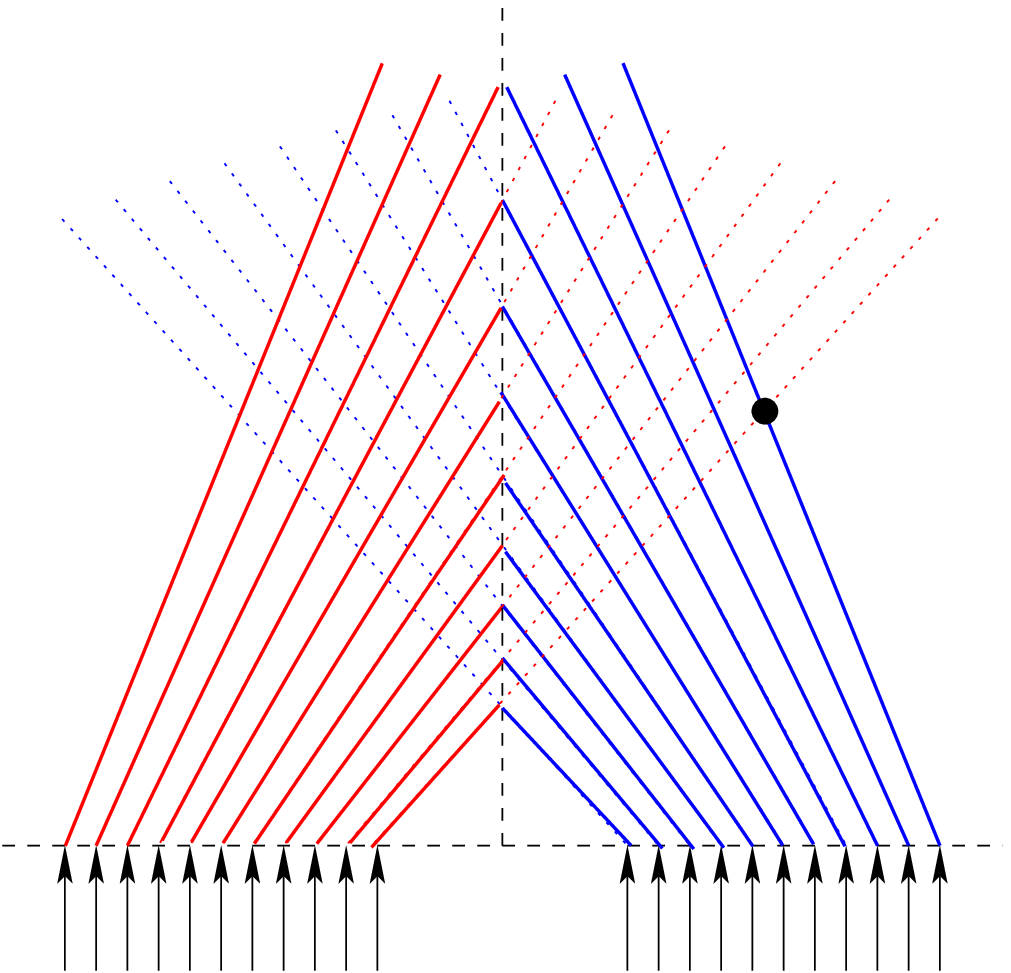}
\begin{picture}(0,0)
\put(-40,90){${\rm ray \ 1}$}
\put(-30,129){${\rm ray \ 2}$}
\put(-116,199){${\rho=0}$}
\put(-46,110){${\mathcal{P}}$}
\put(-155,4){${\rm weak \ shock \ null \ generators}$}
\put(-5,28){$u=0$}
\put(-5,18){${\rm strong \ shock}$}
\end{picture}
\end{center}
\caption{Diagram illustrating (a section of) the \textit{spatial} trajectories of the null generators of the weak shock, exhibiting their focusing after $u=0$. Points at the axis ($\rho=0$) will be hit, simultaneously, by an infinite number of null generators, two of each are in the represented section. After this event, which is at the caustic, spacetime becomes curved therein, and therefore the continuation of the rays beyond the axis is merely illustrative and is represented as a dotted line. For points outside the axis, such as point $\mathcal{P}$, this diagram suggests an initial radiation signal, associated to ray 1, followed by a later burst, associated to ray 2. We shall see this interpretation matches the wave forms computed below.}
\label{rays}
\end{figure}

 The computation will be performed in the more intuitive Brinkmann type coordinates. To obtain the boundary condition we transform back to these coordinates on $u=0^+$. Using \eqref{ct} with $\kappa$ replaced by $\nu$ we find
 \begin{equation}
 g_{\mu\nu}=\frac{\partial \tx^\alpha}{\partial x^\mu} \frac{\partial \tx^\beta}{\partial x^\nu} \bar{g}_{\alpha\beta} \ , \ \ \left.\frac{\partial \tx^\alpha}{\partial x^\mu}\right|_{u=0^+} =\left(
 \begin{array}{ccc}
1 & 0 & 0\\ \frac{\nu^2\Phi'^2}{4} & 1 &  -\frac{\nu \Phi'x^j}{\rho}  \\  -\frac{\nu x^i\Phi'}{2\rho}  & 0 & \delta^{ij}
\end{array} 
\right)  \ ,
\end{equation}
where $\bar{g}_{\alpha\beta}$ is the metric for the weak shock by taking $\tu=0$ in~\eqref{collision}. Observe that the dimension of the parameters $\nu$ and $\lambda$ is $[{\rm Length}]^{D-3}$, so we choose to perform the following rescaling to dimensionless coordinates $(u,v,x^i)\rightarrow \nu^{1/(D-3)} (\sqrt{2}u,\sqrt{2}v,x^i)$.
Performing the matrix multiplication and expressing everything in the unbarred dimensionless coordinates, we write the geometry, on $u=0^+$, as
\begin{equation} g_{\mu\nu}=\nu^{\frac{2}{D-3}}\left[\eta_{\mu\nu}+\frac{\lambda}{\nu}h_{\mu\nu}^{(1)}+\left(\frac{\lambda}{\nu}\right)^2h_{\mu\nu}^{(2)}\right]\ , \label{metric12}\end{equation}
where the flat metric is 
\begin{equation} \eta_{\mu\nu}dx^\mu dx^\nu=-2dudv+\delta_{ij}dx^idx^j \ . \label{background} \end{equation}
 The perturbations\footnote{The metric \eqref{metric12} is exact on $u=0^+$, even though it has been written as a perturbation of flat spacetime.} are
\begin{equation}
\begin{array}{c}
\displaystyle{h_{uu}^{(1)}= (D-3)\Phi'^2h(v,\rho)\ , \qquad  h_{ui}^{(1)}=-\frac{ x_i}{\rho}\sqrt{2}(D-3)\Phi'h(v,\rho)\ ,} \\ \displaystyle{ h_{ij}^{(1)}= 2\left(-\delta_{ij}+(D-2)\frac{x_ix_j}{\rho^2}\right)h(v,\rho)\ ,}
\end{array} \label{fo}
\end{equation}
\begin{equation}
\begin{array}{c}
\displaystyle{h_{uu}^{(2)}= \frac{(D-3)^2}{2}\Phi'^2h(v,\rho)^2 \ , \qquad  h_{ui}^{(2)}=-\frac{ x_i}{\rho}\frac{\sqrt{2}(D-3)^2}{2}\Phi'h(v,\rho)^2\ ,}\\
\displaystyle{ h_{ij}^{(2)}=\left(\delta_{ij}+(D-2)(D-4)\frac{x_ix_j}{\rho^2}\right)h(v,\rho)^2 \ ,}
\end{array}
\end{equation}
where
\begin{equation}
h(v,\rho)\equiv-\dfrac{\Phi'}{2\rho}\left(\sqrt{2}v-\Phi\right)\theta\left(\sqrt{2}v-\Phi\right) \; .
\end{equation}
Observe that the step function jumps at $v=\Phi/\sqrt{2}$. To understand why, recall that, from \eqref{collision}, the collision occurs at $\bar{u}=0=\bar{v}$, in Rosen coordinates. From \eqref{ct}, taking into account the rescaling to dimensionless coordinates performed above, the collision takes place at\[ u=0 \ , \qquad v=\frac{\Phi(\rho)}{\sqrt{2}} \ , \]
in Brinkmann coordinates. The future light cone of the collision has two branches: $\bar{u}=0, \bar{v}>0$ and $\bar{u}>0, \bar{v}=0$. In terms of the Brinkmann coordinates these conditions read:
\[ u=0 \ , \ \  v>\frac{\Phi(\rho)}{\sqrt{2}} \ , \qquad {\rm  and} \qquad u>0 \ , \ \ v=\frac{\Phi(\bar{\rho})}{\sqrt{2}}+u\frac{\Phi'(\bar{\rho})^2}{4} \ , \  \ {\rm where} \ \ \rho=\bar{\rho}\left(1+\frac{u\Phi'(\bar{\rho})}{\sqrt{2}\bar{\rho}}\right) \ . \]
In order to understand the interaction, let us follow the null generators of the shock with support at $\bar{v}=0$ (which we call the \textit{weak} shock). Before the collision these can be parametrised as
\[ \bar{u}=\Lambda \ , \qquad \bar{v}=0 \ , \qquad \bar{x}^i=\xi^i \ \Rightarrow \bar{\rho}=\sqrt{\xi^i\xi^j\delta_{ij}}\equiv \xi \  \ . \]
To go beyond the collision we look at Brinkmann coordinates. From \eqref{ct}:
\begin{eqnarray}
u &=& \Lambda\ ,\qquad 
v = \theta(\Lambda)\left(\frac{\Phi(\xi)}{\sqrt{2}} + \frac{\Lambda \Phi'(\xi)^2}{4}\right)\ , \qquad x^i= \xi^i\left(1-\frac{\sqrt{2}\Lambda \theta(\Lambda)}{\xi^{D-2}}\right)
\ . \label{ct2}
\end{eqnarray}
In terms of these coordinates the focusing effect on the weak shock null generators as they cross $u=0$ is clear - Fig. \ref{focusing4}, \ref{focusing} and \ref{rays}. Observe that there is a qualitative difference between the behaviour in $D=4$ and $D>4$: the discontinuous jump in $v$  of the weak shock null generators is always positive in $D>4$ but it becomes negative for large $\rho$ in $D=4$. This will give rise to a different domain in the time integration when computing the overall energy emitted. In all cases, nevertheless, the generators will focus, generating a caustic, at $\sqrt{2}\Lambda=\xi^{D-2}$, corresponding to 
 ~
\begin{equation}
\rho=0 \ , \qquad \sqrt{2}v=\left\{ \begin{array}{c} -\ln(\sqrt{2}u)+1 \ , \qquad D=4 \\
\displaystyle{\frac{D-2}{D-4}\frac{1}{(\sqrt{2}u)^{\frac{D-4}{D-2}}} \ , \qquad D>4}   \end{array} \right.  \ .
\label{caustic}
\end{equation}
After crossing at the caustic the generators enter the curved region of the spacetime. Thus, until they cross, the light cone structure is that of the Minkowski spacetime \eqref{background}. It is simple to see that the deflection angle $\alpha$ undergone by a null shock generator at distance $\rho$ from the axis obeys, in Brinkmann coordinates, 
\begin{equation}
\tan\alpha=\frac{\sqrt{2}}{\rho^{D-3}} \ . 
\end{equation}
Thus the focusing increases (decreases) with $D$ for short (long) distances, as expected from the behaviour of the gravitational force.

\subsection{Future development of the metric}
\begin{figure}[t]
\begin{center}
\begin{picture}(0,0)
\put(100,120){$u$}
\put(217,57){$(0,v',\vec{x}')$}
\put(260,44){$\vec{x}$}
\put(266,120){$v$}
\put(194,157){$(u,v,\vec{x})$}
\put(323,104){$u=0$}
\put(323,94){hypersurface}
\end{picture}
\includegraphics[scale=0.4]{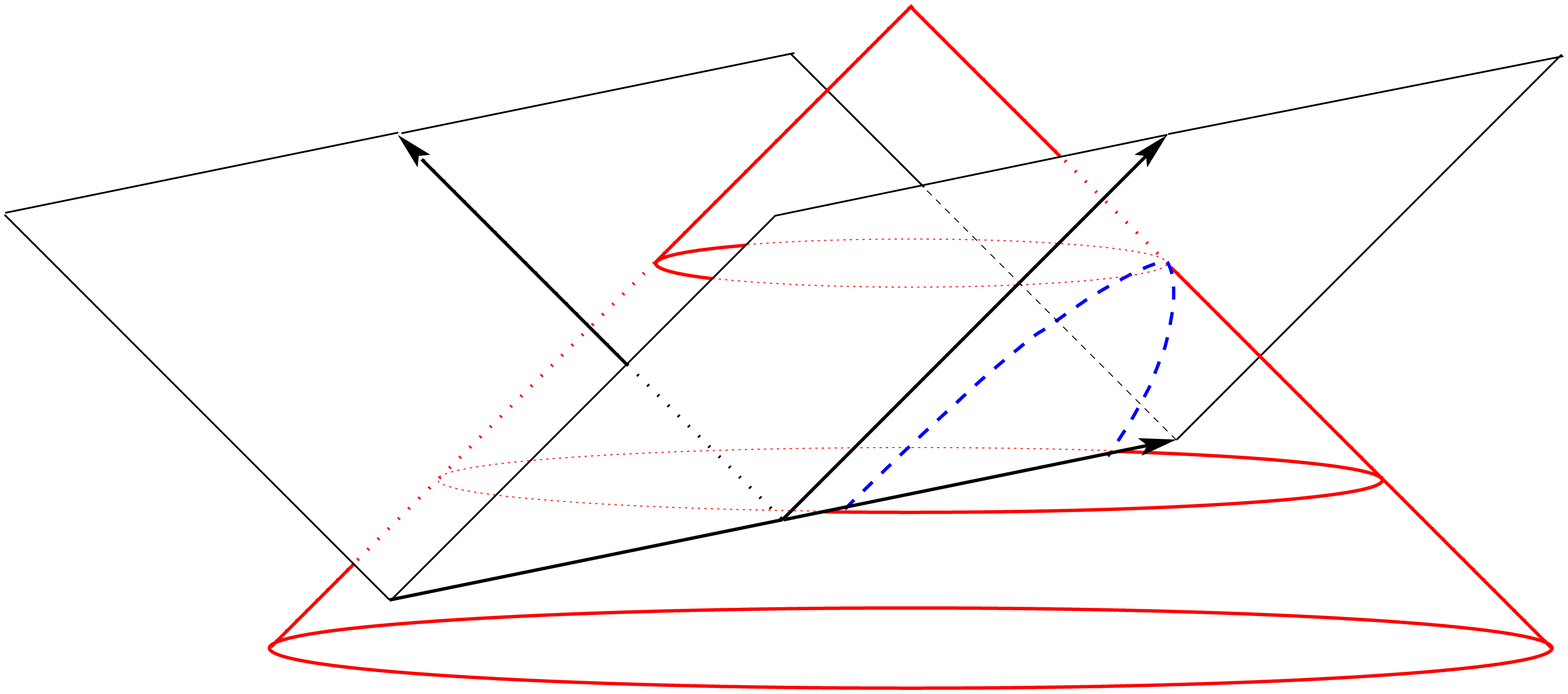}
\end{center}
\caption{The intersection of the past light cone of a spacetime event $(u,v,\vec{x})$, to the future of $u=0=v$, with the null hypersurface $u=0$, is a parabola (dashed blue curve); its points $(0,v',\vec{x}')$ obey \eqref{parabola}. This diagram represents the causal structure of the background metric for unspecified initial data.}
\label{pastlightcone}
\end{figure}

To the future of $u=0$ the metric will have a perturbative expansion of the type
\begin{equation} g_{\mu\nu}=\nu^{\frac{2}{D-3}}\left[\eta_{\mu\nu}+\sum_{i=1}^\infty\left(\frac{\lambda}{\nu}\right)^i h_{\mu\nu}^{(i)}\right]\ ,\label{eq:pertexpansion} \end{equation}
and we should be able to find $h_{\mu\nu}^{(i)}$ by successively solving the Einstein equations at each order in $\lambda/\nu$, with the boundary condition expressed before. To first order we therefore have to solve the linearised $D$-dimensional Einstein equations around Minkowski space, so in the remainder we consider the linear perturbation $h_{\mu\nu}$ in~\eqref{eq:pertexpansion} and for notational simplicity drop the ``(1)'' label. We denote the trace reversed perturbation by barring it $\bar{h}_{\mu\nu}={h}_{\mu\nu}-\eta_{\mu\nu}{h}/2$, and perform a coordinate transformation of the form $x^{\mu}\rightarrow x^{N\mu}+(\lambda/\nu)\xi^{\mu}$
\begin{equation}
h_{\mu\nu}\rightarrow h^N_{\mu\nu}=h_{\mu\nu}+2\xi_{(\mu,\nu)} \ . \label{newper}
\end{equation}
 This can be chosen such that the De Donder gauge condition is obeyed in $u>0$
\begin{equation}
{\bar{h}^{N \,\alpha\beta}}_{\ \ \  \ \ ,\alpha}=0 \ , \label{gauge}
\end{equation}
so that the linearised Einstein equations become simply a set of decoupled wave equations in Minkowski space for each component:
\begin{equation}
\Box \bar{h}^N_{\mu\nu}=0 \ \  \Leftrightarrow \ \ \left(-2\partial_u\partial_v+\partial_i^2\right)\bar{h}^N_{\mu\nu}=0 \ . \label{hareq}
\end{equation}
In~\cite{D'Eath:1992hb}, and integral solution for the wave equation $\Box F=0$ in $u\geq 0$ with initial data on $u=0$ was used. In appendix~\ref{apa}, we show that in $D$ dimensions, the corresponding solution is 
\begin{equation}
F(u,v,x_i)=\dfrac{1}{(2\pi u)^{\frac{D-2}{2}}}\int d^{D-2}x'\partial_{v'}^{\frac{D-2}{2}} F(0,v',x_i')  \; , \label{eq:integral_sol}
\end{equation}
where, for each $x'$, $v'$ defines points at the $u=0$ hypersurface, which are \textit{on} the past light cone of the event $(u,v,x_i)$ (cf. Fig. \ref{pastlightcone}):
\begin{equation}
v'=v-\dfrac{|x-x'|^2}{2u} \; , \label{parabola}
\end{equation}
and the derivative operator is suitably defined in Fourier space (with respect to $v$) for odd dimensions (see appendix  \ref{apa}).
Similarly to~\cite{D'Eath:1992hb}, the gauge condition~\eqref{gauge} is obeyed in $u>0$ if 
\begin{equation}
{\bar{h}^{N \,\beta}}_{ \ \ \ \alpha,\beta v} \big|_{u=0}=0 \ .  \label{gauge2}
\end{equation} 
To see this we first contract the wave equation with another derivative to get $\Box \bar{h}^{N\ , \nu}_{\mu\nu}=0$. It then follows from  the integral solution \eqref{eq:integral_sol}, that if
\begin{eqnarray} \left.\partial^{\frac{D-2}{2}}_v\th^{N\, \ \beta}_{\ \ \ \ \ \alpha  \ ,\beta } \right|_{u=0}&=&0 \ ,  \label{gauge3}
\end{eqnarray}
then the De Donder condition holds everywhere. In particular~\eqref{gauge2} is sufficient, since it implies~\eqref{gauge3}.

\subsection{Gauge fixing at $u=0$}
Condition~\eqref{gauge2} can be written, in terms of the perturbations in the original coordinate system, using \eqref{newper} (on $u=0$). Then, using~\eqref{hareq} and the fact that the trace of the first order metric perturbation $h$ vanishes on $u=0$, we find, that~\eqref{gauge2} implies, on $u=0$
\begin{eqnarray}
\hspace{-0.9cm} \frac{1}{2}\Box\left(\xi_{i,}^{\ \,  i} -2\xi_{u,v}\right)&=&h^{\ \ \ i}_{u i, \ v}+h^{\ \ \ u}_{u u, \ v}-\frac{1}{2}h^{\ \ \ i}_{u v, \ i}=-(D-3)(D-2)\left[\dfrac{\Phi'}{\rho}\right]^2\theta\left(\sqrt{2}v-\Phi\right)  ,  \label{per1} \\
-\Box\xi_{[j,v]}&=&h^{\ \ \ i}_{j i, \ v}+h^{\ \ \ u}_{j u, \ v}-\frac{1}{2}h^{\ \ \ i}_{j v, \ i} \ =0 \; , \label{per2} \end{eqnarray}
where the right hand side  of \eqref{per1} and \eqref{per2} was computed using \eqref{fo}.
We look for a solution of \eqref{per1} and \eqref{per2} which has a power series expansion around $u=0$:
\begin{equation}
\xi_\mu(u,v,x^i)=\xi_\mu^{(0)}(v,x^i)+u\xi_\mu^{(1)}(v,x^i)+\dots 
\end{equation}
One such solution is $\xi_\mu^{(0)}=\xi_v^{(1)}=\xi_{i}^{(1)}=0$ and
\begin{eqnarray}
 \xi_u^{(1)}&=&-\dfrac{(D-3)(D-2)}{8}\left(\dfrac{\Phi'}{\rho}\right)^2\left(\sqrt{2}v-\Phi\right)^2\theta\left(\sqrt{2}v-\Phi\right) \ . 
\end{eqnarray}
Applying this gauge transformation, only the $uu$ first order perturbation changes in this new (De Donder) gauge, on $u=0$, which becomes 
\begin{equation}
h_{uu}^{N}= (D-3)\Phi'^2h(v,\rho)+2\xi_u^{(1)} \ . \label{fo2}
\end{equation}
Since the metric perturbation is traceless on $u=0$, it is traceless everywhere, due to~\eqref{eq:integral_sol} (because $\partial_v h^N(0,v,x^i)=0$ so the solution is identically zero). Thus~\eqref{hareq} becomes
\begin{equation}
\Box h^{N}_{\mu\nu}=0 \ \  \Leftrightarrow \ \ \left(-2\partial_u\partial_v+\partial_i^2\right) h^{N}_{\mu\nu}=0 \ . \label{hareq2}
\end{equation}
The general solution of~\eqref{hareq2} in $u>0$ is again obtained using~\eqref{eq:integral_sol} with the replacement $F\rightarrow h^{N}_{\mu\nu}$.
Observe that the components of the metric perturbation that vanish at $u=0$, will vanish when $u>0$. We can boost back to the centre of mass frame simply by using the replacements $\{u,v\}\rightarrow \{e^{-\alpha}u,e^{\alpha}v\}$.  At linear order, we can use $x_\mu^N=x_{\mu}$ for the coordinates in the new gauge. Finally, we can change from units of $\nu^{\frac{1}{D-3}}$ to units of $\kappa^{\frac{1}{D-3}}$ and obtain the linearised, centre of mass frame metric in $u>0$ 
\begin{equation}
\dfrac{ds^2}{\kappa^{\frac{2}{D-3}}}=-2dudv+h^{N}_{uu}(u,v,x_k)du^2+2h^{N}_{ui}(u,v,x_k)dudx^i+\left[\delta_{ij}+h^{N}_{ij}(u,v,x_k)\right]dx^idx^j \; .
\end{equation}

\section{Extracting the gravitational radiation}
\label{section4}
To extract the gravitational radiation produced in the collision we shall use the Landau-Lifshitz pseudo-tensor \cite{landau}, which was generalised to higher dimensions in \cite{Yoshino:2009xp}. In terms of our perturbations $h^{N}_{\mu\nu}(u,v,x_i)$, taking into account that they are traceless, it reads (we omit from now on the superscript $N$ for notational simplicity) \cite{Yoshino:2009xp}:
\begin{equation}
\begin{array}{rcl}
16\pi G_Dt^{\mu\nu}_{LL} & = & \displaystyle{h^{\mu\nu}_{\ \ , \alpha}h^{\alpha\beta}_{\ \ , \beta}-h^{\mu\alpha}_{\ \ , \alpha}h^{\nu\beta}_{\ \ , \beta}+\frac{1}{2}\eta^{\mu\nu}\left(h^{\alpha\beta}_{\ \ , \sigma}h^\sigma_{\ \alpha,\beta}-\frac{1}{2}h^{\beta\sigma,\alpha}h_{\beta\sigma,\alpha}\right)} \\
& & \displaystyle{-h^{\mu\beta}_{\ \ , \sigma}h_{\beta}^{\  \sigma, \nu}-h^{\nu\beta}_{\ \ , \sigma}h_{\beta}^{\  \sigma, \mu}+h^{\mu\alpha,\beta}h^\nu_{\ \alpha,\beta}+\frac{1}{2}h^{\beta\sigma,\mu}h_{\beta\sigma}^{\ \ ,\nu}} \ . \label{LL}
\end{array}
\end{equation}
Despite not being unique or gauge-invariant it is well known that the integral
\begin{equation}
E_{\rm radiated}=\int t^{0i}_{LL}n_idSdt \ , 
\end{equation}
computed on a `distant' surface with area element $dS$ outward unit normal $n^i$, is a gauge-invariant well defined energy \cite{wald}.

Following~\cite{D'Eath:1992hb}, we shall compute the power emitted along the vicinity of the $\theta=\pi$ direction (in the CM frame), corresponding to the negative $z$ direction. This maps to the region where perturbation theory should be valid in the boosted frame (cf. Fig. \ref{spacetime}). So we shall need the flux along the $z$ direction which is given by ($t=(v+u)/\sqrt{2}$ and $z=(v-u)/\sqrt{2}$):
\begin{equation}
t_{LL}^{0z}=\frac{1}{2}\left(t^{vv}_{LL}-t^{uu}_{LL}\right) \ . \label{t0z}
\end{equation}
The first (second) term corresponds to the flux across a $v={\rm constant}$ ($u={\rm constant}$) surface. Thus, in order to determine the flux in the $\theta=\pi$ direction we need only the second term. Then, from \eqref{LL} (observe that the first two terms are zero by the De Donder gauge condition):
\begin{equation}
t_{LL}^{0z}\big|_{\theta=\pi}=-\frac{1}{2}t^{uu}_{LL}\big|_{\theta=\pi}=-\frac{1}{64\pi G_D}h^{ij}_{\ \ ,v}h_{ij,v}\big|_{\theta=\pi}  \ .
\end{equation}
The total radiation emitted will be computed, to first order, under the assumption that $dE/d\cos\theta$ is isotropic (we shall further discuss this approximation below). Thus we integrate the power emitted inside the narrow cone around the $\hat{\theta}\equiv\pi-\theta=0$ axis. Using $dS= r^{D-2}d\Omega_{D-2}$, taking the limit close to the axis and multiplying by the area of the sphere of radius $r$, this energy is 
\begin{equation}
E_{\rm radiated}=  \frac{\Omega_{D-3}}{32\pi G_D}\lim_{\hat{\theta}\rightarrow 0,r\rightarrow \infty}\left(r^2\rho^{D-4}\int h^{ij}_{\ \ ,v}h_{ij,v} dt \right)\ .
\end{equation}
As expected, only the $ij$ components (which did not change with our gauge choice) determine the energy emitted in gravitational radiation. In appendix~\ref{apb} we show that, in fact, there is only one independent quantity that determines the integrand in the last equation, denoted $E=E(u,v,\rho)$.
In terms of $E$, we can express the radiated energy as
\begin{equation}\label{rad_dimensionfull}
E_{\rm radiated}=  \frac{\Omega_{D-3}}{32\pi G_D}\dfrac{D-2}{D-3}\lim_{\hat{\theta}\rightarrow 0,r\rightarrow \infty}\left(r^2\rho^{D-4}\int (E_{,v})^2 dt \right) \equiv \epsilon_{\rm radiated} 2\mu \; .
\end{equation}
We identify, inside the parenthesis, the analogous of Bondi's news function used in~\cite{D'Eath:1992hb}. The coordinates used in the last formula are dimensionful.
The fraction of the energy radiated in gravitational waves  $\epsilon_{\rm radiated}$ defined in~\eqref{rad_dimensionfull}  is given by,
\begin{equation}
\epsilon_{\rm radiated}=  \frac{1}{8}\dfrac{D-2}{D-3}\lim_{\hat{\theta}\rightarrow 0,r\rightarrow \infty}\left(\int (r\rho^{\frac{D-4}{2}}E_{,v})^2 dt \right) \; , \label{rad1}
\end{equation}
and, from appendix~\ref{apb}
\begin{equation}
 E_{,v}=-\dfrac{\sqrt{8}\Omega_{D-4}}{(2\pi u)^{\frac{D-2}{2}}} \int_{0}^{+\infty} \dfrac{d\rho'}{\rho'}\,\int_{-1}^{1} dx \,\dfrac{d}{dx}\left[x(1-x^2)^{\frac{D-3}{2}}\right] \delta^{(\frac{D-4}{2})}\left(v_1'-v_2'\right) \label{rad2}
 \; ,
\end{equation}
where 
\begin{equation}
v_1'\equiv v-\frac{\rho^2-2\rho\rho'x+\rho'^2}{2u} \ ,\qquad \ v_2'\equiv \frac{\Phi(\rho')}{\sqrt{2}} \ .
\label{v1v2}
\end{equation}
All coordinates in \eqref{rad1} and \eqref{rad2} are now the dimensionless coordinates used in Section \ref{sec3}. 

Observe that at $\rho=0$, the argument of the delta function has no $x$ dependence and the $x$ integral can be immediately performed to yield zero. This can be interpreted as the result of a destructive interference phenomenon. To see this consider for instance the $h_{11}$ component of the initial data \eqref{fo}, which can be written as 
\begin{equation}
h_{11}=2h(v',\rho')\left\{\begin{array}{ll} 2x^2-1 \ , & D=4 \vspace{2mm} \\ \displaystyle{\frac{2}{D-4}C_2^{\left(\frac{D-4}{2}\right)}(x)} \ , & D>4 \end{array} \right. \ .
\end{equation}
The angular part of the perturbation is a scalar harmonic with $\ell=2$ (or $\ell=-2$ in $D=4$) on the $(D-3)$-sphere. Integrating such scalar harmonic on the $(D-3)$-sphere gives zero. This is the integral appearing in \eqref{rad2} at the axis and may be seen as a cancellation between the different phases coming from different points on the integration circle of radius $\rho'$.

The vanishing of the power radiated at the axis in a head on collision of two equal objects (like particles or black holes) can also be physically interpreted as follows. Gravitational radiation emission is determined by the variation of the gravitational quadrupole. Taking the observation point at the collision axis, and behind one of the objects, no quadrupole variation is observed.\footnote{We thank U. Sperhake for this observation.}

\subsection{Integration limits}
We now discuss the integration domain for the time integration in \eqref{rad1}. Equation \eqref{rad2} suggests that the gravitational radiation seen at a spacetime point $\mathcal{P}$ to the future of the collision, originates from the points on the collision surface wherein the delta function has support; these are the points at  the intersection of the past light cone of $\mathcal{P}$ with the collision surface, as expected.\footnote{Observe, however, that an integration in $v'$ was already performed to get \eqref{rad2}. For even $D$ this integration is performed using a delta function which enforces the support of the integrand on the light cone. But for odd $D$ the initial integral has support also inside the light cone, cf. Appendix \ref{apa}.} 

\begin{figure}[t]
\begin{center}
\begin{picture}(0,0)
\put(144,101){$u$}
\put(311,99){$v$}
\put(377,125){$_{u=0}$}
\put(290,76){$\frac{1}{\rho'^{D-4}}$}
\put(119,109){$_{\tau=0}$}
\put(307,39){$\vec{x}$}
\put(117,155){$_{\tau=\tau_1}$}
\put(139,170){$\tau$}
\put(187,175){$_{z=-|z|= {\rm constant}}$}
\put(275,165){$_{\rm caustic}$}
\end{picture}
\includegraphics[scale=0.4]{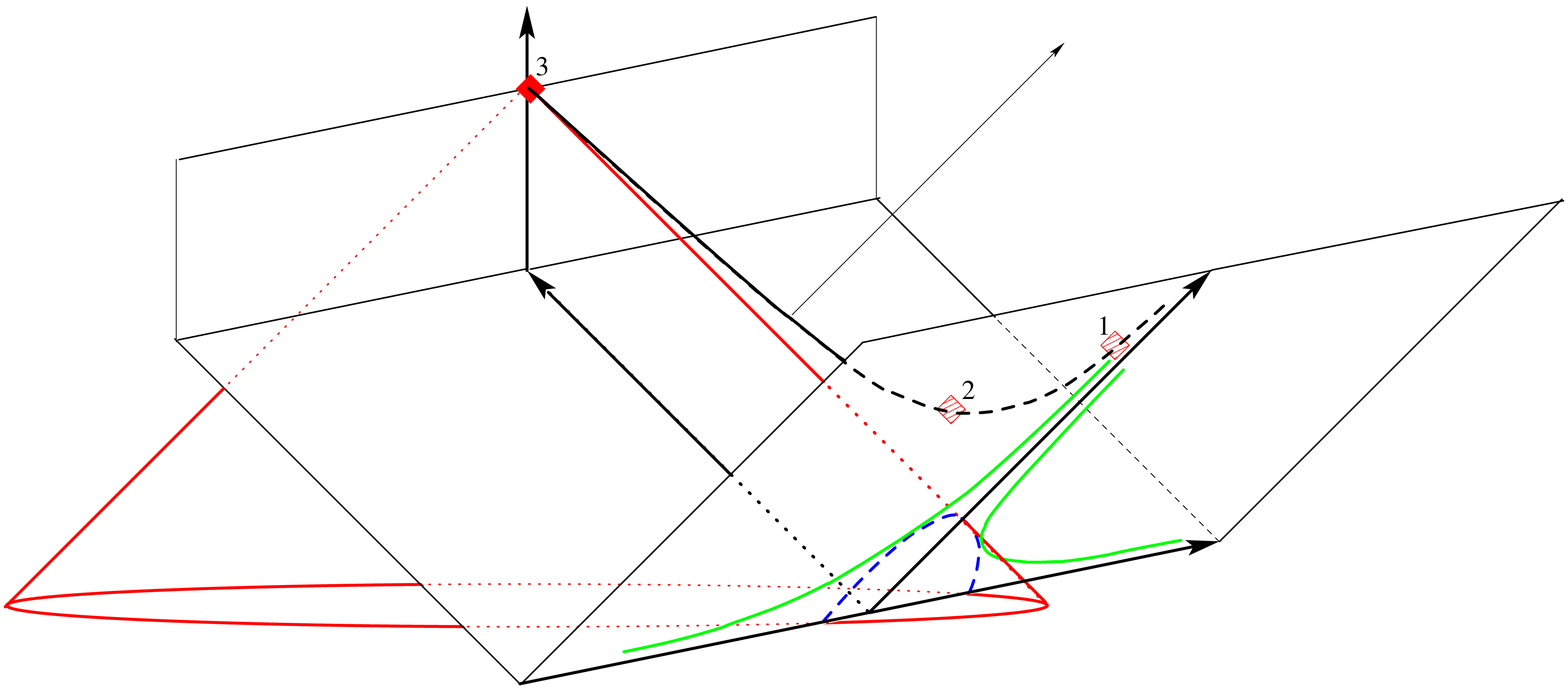}
\end{center}
\caption{The caustic is represented in a spacetime diagram for the case $D>4$. Along the caustic three events $1-3$ are selected. For event $3$ its past light cone is drawn and its intersection with the $u=0$ surface is represented as the blue dashed line (parabola). This is tangent to the collision surface $u=0$ and $v=\Phi/\sqrt{2}$, represented by the solid (green) lines. The generators of the shock traveling along $u$ that emerge from the intersection points will focus and converge at $3$. For an observer at fixed $z$, following the worldline represented by the $\tau$ axis, no gravitational radiation will be observed before $\tau=\tau_1$. In Fig. \ref{tangent} (right panel) the collision lines and the intersections with $u=0$ of the past light cones of other points along the caustic are represented.}
\label{pastlightcone3}
\end{figure}

\begin{figure}[t]
\begin{center}
\includegraphics[scale=0.3]{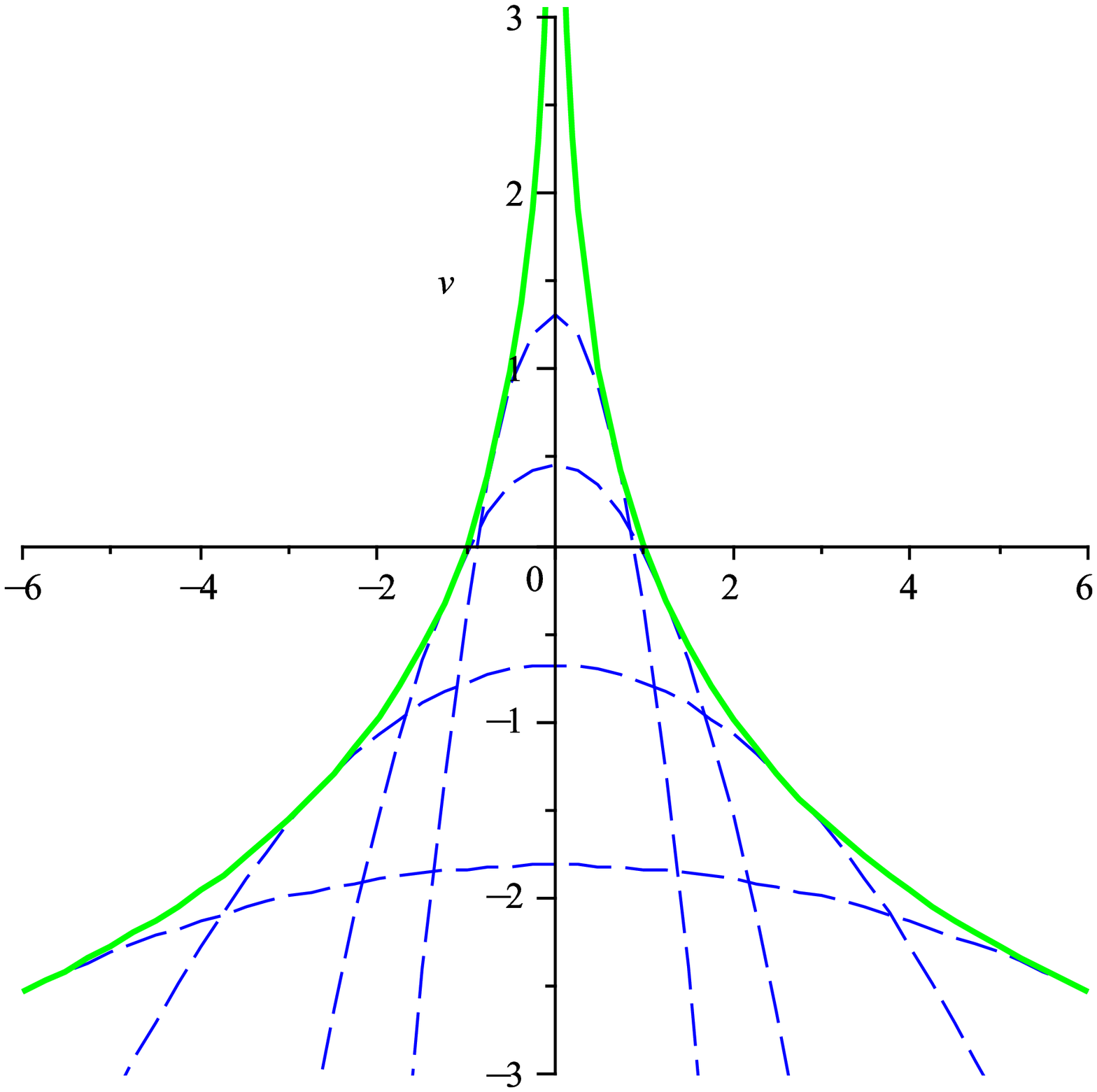} \hspace{1cm}
\includegraphics[scale=0.3]{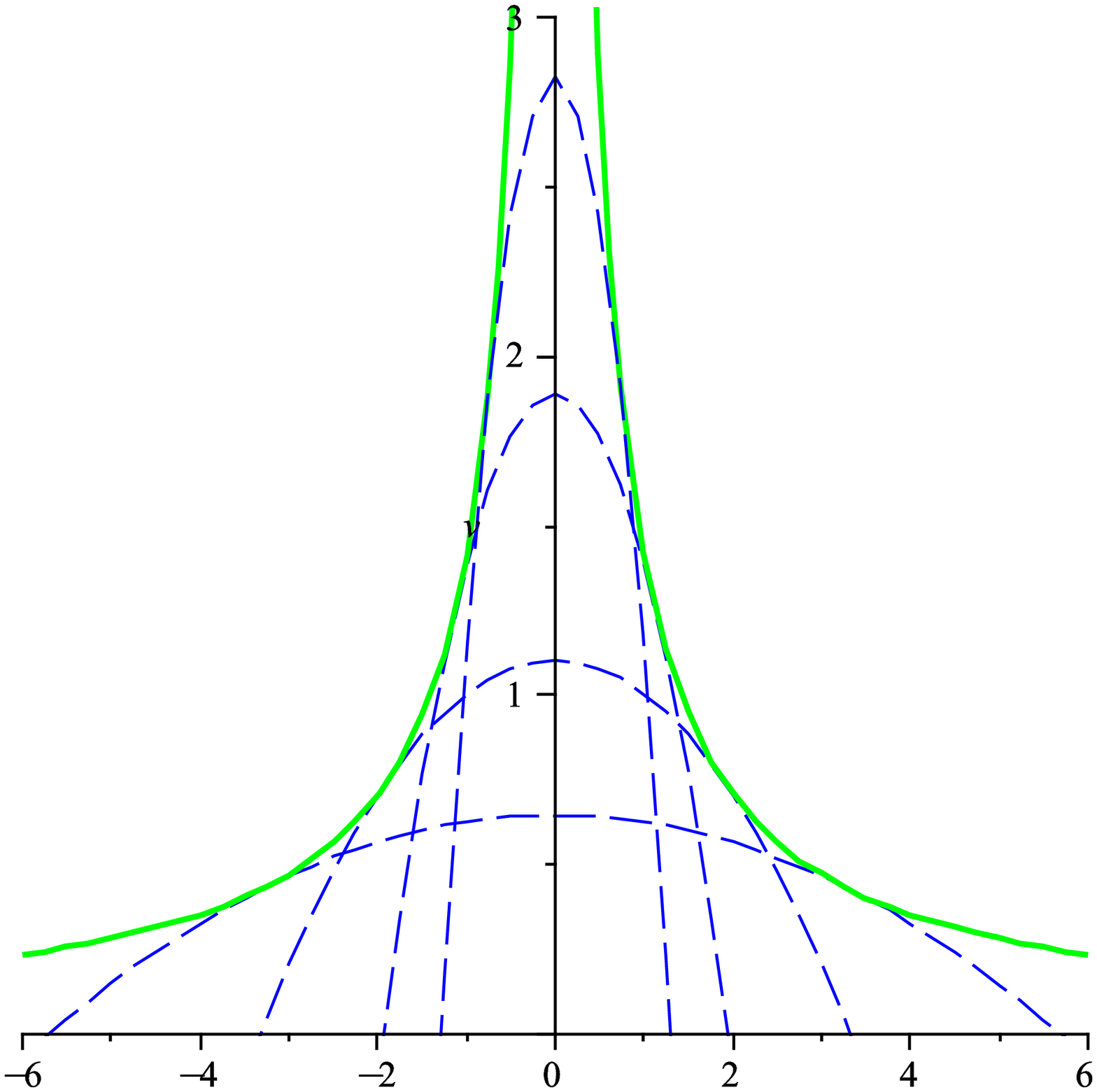}
\begin{picture}(0,0)
\put(-85,155){$_{\bf 1}$}
\put(-85,109){$_{\bf 2}$}
\put(-85,72){$_{\bf 3}$}
\end{picture}
\end{center}
\caption{Collision surface (at $u=0$ and $v=\Phi/\sqrt{2}$ - green solid lines) and the past light cones at $u=0$ (blue dashed lines) of various spacetime points along the caustic for $D=4$ (left panel) and $D=5$ (right panel). The past light cones are tangent to the collision surface. The latter case is qualitatively similar to any $D>5$. The numbering 1-3 in the right panel corresponds to the spacetime points with similar numbering in Fig. \ref{pastlightcone3}.}
\label{tangent}
\end{figure}

\begin{figure}[t]
\begin{center}
\includegraphics[scale=0.3]{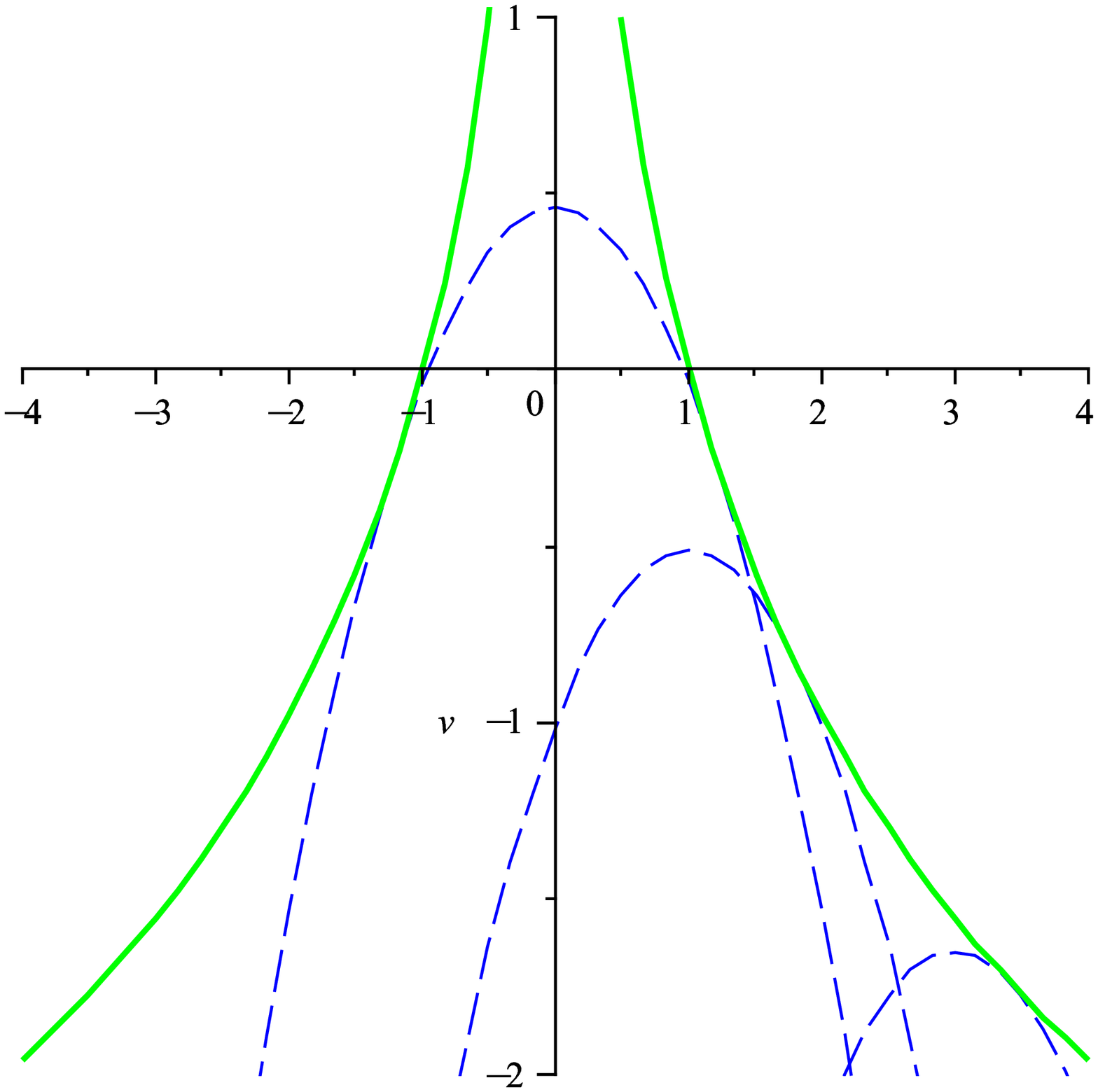} \hspace{1cm}
\includegraphics[scale=0.3]{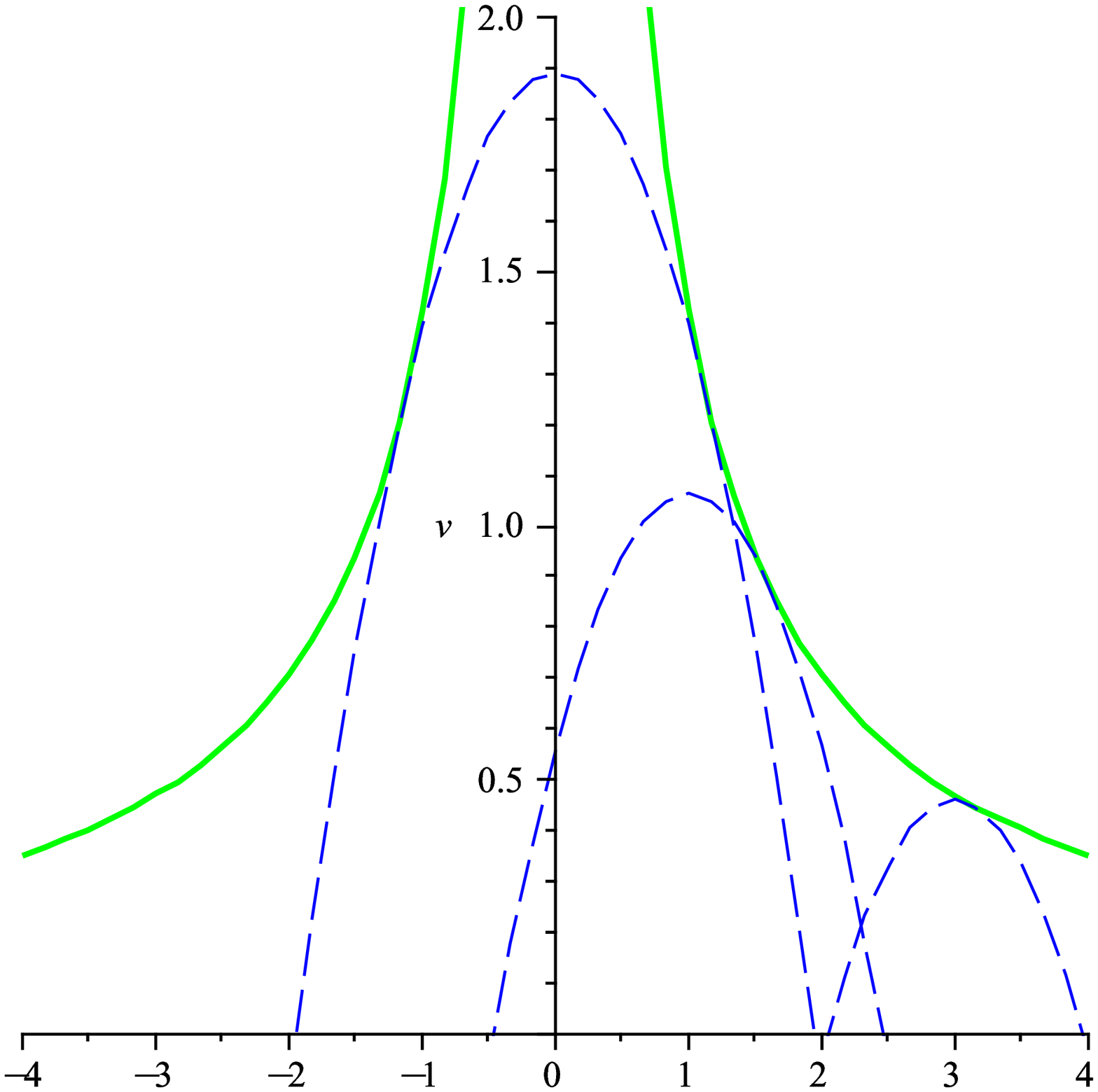}
\begin{picture}(0,0)
\end{picture}
\end{center}
\caption{Same as the previous figure but now for spacetime points away from the axis (with the same $u=1$ but different distances from the axis $\rho=0,1,3$) such that their past light cone at $u=0$ is tangent to the collisions surface, for $D=4$ (left panel) and $D=5$ (right panel). The earliest intersection always occurs for $x=1$.}
\label{tangentnotaxis}
\end{figure}

Instead of using the spacetime coordinates $(u,v,\rho)$ it is convenient to introduce a retarded time coordinate $\tau=t-r$, where the radial coordinate is $r=\sqrt{z^2+\rho^2}$ and an angular coordinate $\theta$ by $\rho=r\sin\theta$. Thus we wish to know the first instant $\tau_1$ for which the point specified by $(r,\theta)$ receives gravitational radiation. This event occurs when the intersection of its past light cone with $u=0$ becomes tangent to the collision surface and is determined in a similar way to the determination of the caustic in Section \ref{sec3}. One finds that $\tau_1=\tau_1(r,\theta)$, together with the auxiliary variable $\bar{\rho}$, is obtained by solving the system of equations
\begin{equation}
\begin{array}{c}
\displaystyle{r\sin\theta=s\bar{\rho}\left(1-\frac{\tau+2r\sin^2\frac{\theta}{2}}{\bar{\rho}^{D-2}}\right)} \vspace{0.3mm} \\
\displaystyle{\tau\left(1-\frac{1}{\bar{\rho}^{2D-6}}\right)+2r\left(\cos^2\frac{\theta}{2}-\frac{\sin^2\frac{\theta}{2}}{\bar{\rho}^{2D-6}}\right)=\Phi(\bar{\rho}) }
\end{array} \ , \label{tau12}
\end{equation}
for specified $\rho$ and $\theta$ and $s=+1$. The intersection of the past light cone of points $(\tau_1,r,\theta)$ (points 1 to 3 in Fig.~\ref{pastlightcone3}) with $u=0$ is plotted in Fig. \ref{tangent}, for $\theta=\pi$ and Fig. \ref{tangentnotaxis}, outside the axis. As claimed these curves are tangent to the collision surface.
~

Thus the initial time for observing gravitational radiation, $\tau_1$, is positive, for $D>4$ and goes to zero as $r$ goes to infinity (cf. Fig. \ref{focusing}). By contrast, in $D=4$, $\tau_1$ becomes negative and approaches negative infinity as  $r$ goes to infinity (cf. Fig. \ref{focusing4}). Thus, the time integration will have different domains in $D=4$ and $D>4$.  

The previous computation of $\tau_1$ amounts to computing the retarded time at which ray 1 intersects point $\mathcal{P}$ in Fig. \ref{rays}. The retarded time, $\tau_2$, for which ray 2 intersects $\mathcal{P}$ is computed by solving \eqref{tau12} with $s=-1$. This coincides with a second peak in the wave forms exhibited in the next section.

\subsection{Numerical evaluation}
In this section we finally obtain the wave forms for several dimensions and integrate the radiated power to estimate the amount of gravitational radiation emitted in the collision. First we perform the angular integral in~\eqref{rad2} (see appendix~\ref{apb}  where the polynomials of degree $M+2$,  $P^{(M+2)}$ and  $Q^{(M+2)}$ are defined), to obtain
\begin{equation}
r\rho^{M+\frac{1}{2}}E_{,v}=\dfrac{(-1)^M4\Omega_{D-4}}{(2\pi)^{M+2}(D-1)}\frac{r}{\rho} \int_{\mathcal{D}} \dfrac{d\rho'}{\rho'^{M+\frac{5}{2}}}\dfrac{P^{(M+2)}(x_\star)}{\sqrt{1-x_\star}}
 \, , \label{odd}
\end{equation}
in odd dimensions and 
\begin{equation}
 r \rho^{M}E_{,v}=\dfrac{(-1)^{M}2\sqrt{2}\Omega_{D-4}}{(2\pi)^{M+1}(D-1)}\frac{r}{\rho} \int_{\mathcal{D}} \dfrac{d\rho'}{\rho'^{M+2}}\,\dfrac{ Q^{(M+2)}(x_\star) }{\sqrt{1-x_\star^2}} 
 \; , \label{even}
\end{equation}
in even dimensions. $\mathcal{D}$ is defined such that
\begin{equation}
-1 \leq x_\star \equiv \dfrac{U\Phi\left(\rho'\right)+\rho'^2-UT}{2\rho \rho'} \leq 1\; ,
\label{eq:domain}
\end{equation}
where $U = \tau + 2r\sin^2(\theta/2)$, $T=\tau +2r\cos^2(\theta/2)-\rho^2/U$ and we are now expressing the result in coordinates $\left\{\tau,r,\theta\right\}$ as in Eq.~\eqref{tau12}.

In the remaining numerical analysis, we present only results for even $D$. Concerning odd $D$, we have attempted to evaluate~\eqref{odd}; the wave forms obtained (numerically), however, contained non-integrable tails when squared to get the radiated power. Our best hint, at the moment, is that these divergences are similar to those present in the second order computation \cite{D'Eath:1992hd}. Nevertheless, our aim is to investigate the variation of the radiated energy with $D$. Thus the even case already shows the general trend. It seems plausible that the result for $D$ odd will interpolate between the even $D$ cases we shall exhibit.

To evaluate~\eqref{even}, we wrote firstly a test code in \textsc{mathematica}7 and then a code in the \textsc{c++} language, using the numerical integration routines of the Gnu Standard Library (GSL). The purpose was to check the two codes in some cases and use the (lower level) \textsc{c++} code for generating all the data in a practical amount of time. The domain of integration $\mathcal{D}$ was determined by looking at the roots of two polynomials constructed from $x_\star$ and using the root finding routines in \textsc{mathematica} or the GSL library. In general, depending on the spacetime point, we found either an empty integration domain (giving a value of exactly zero to the wave form), or a non-empty domain which can be simply connected or the union of two disconnected domains. The integrable singularities at the points $x_\star=\pm 1$ where removed explicitly through a change of variable around each singularity. In general, for the wave forms, we have demanded a relative error of $10^{-8}$, and $10^{-5}$ for the final $\tau$ integral of the radiated power.

\begin{figure}[t]
\hspace{-2mm}\includegraphics[scale=0.66,clip=true,trim= 0 0 0 0]{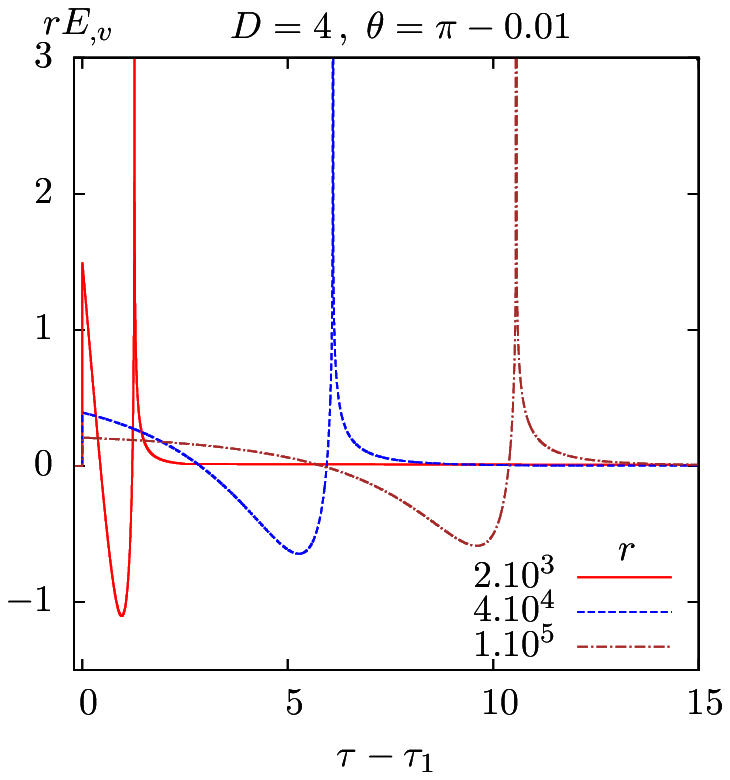} \hspace{1.5mm} 
\includegraphics[scale=0.66,clip=true,trim= 0 0 0 0]{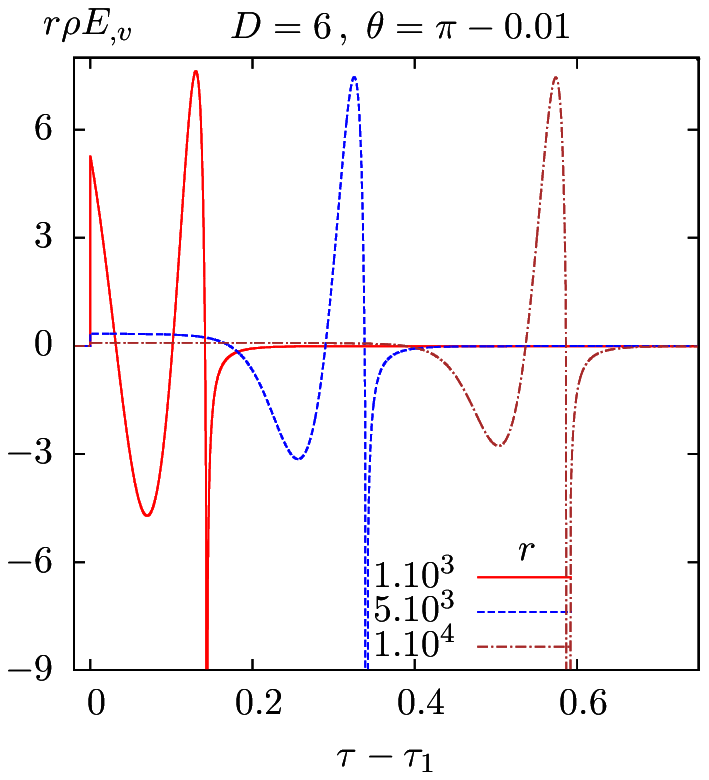} \hspace{1.5mm} 
\includegraphics[scale=0.66,clip=true,trim= 0 0 0 0]{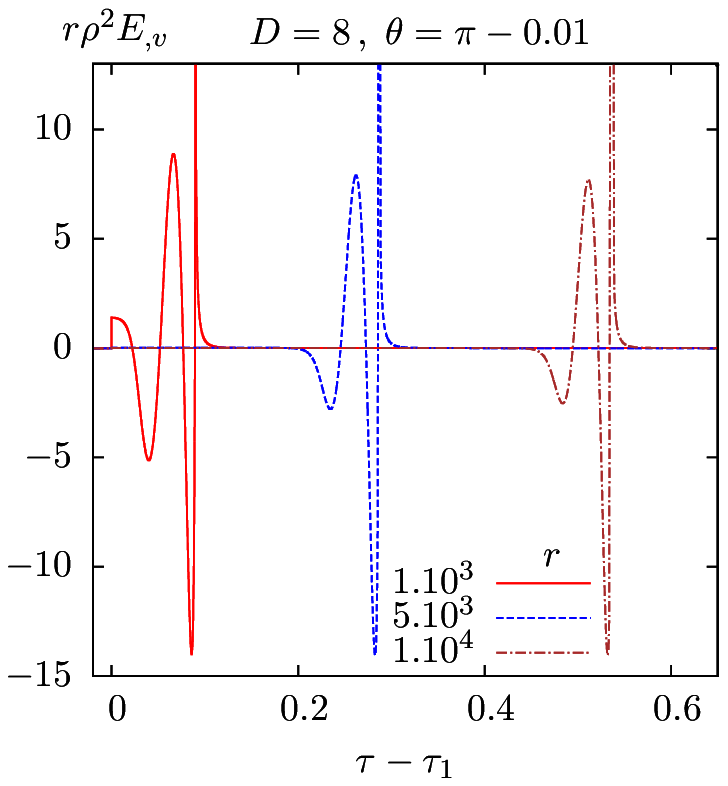}
\caption{\label{fig:rVarDvar} {\em Wave forms:} These plots show the even $D$ wave forms~\eqref{even} for fixed $\theta$, close to the axis of symmetry and for various $r$. Note the initial step which is due to the impulsive nature of the initial conditions associated with the Aichelburg-Sexl shocks. This step is exactly at $\tau=\tau_1$, whereas the sharp peak is at $\tau=\tau_2$, for each curve.}
\end{figure}

\begin{figure}[t]
\hspace{-3mm}\includegraphics[scale=0.66,clip=true,trim= 0 0 0 0]{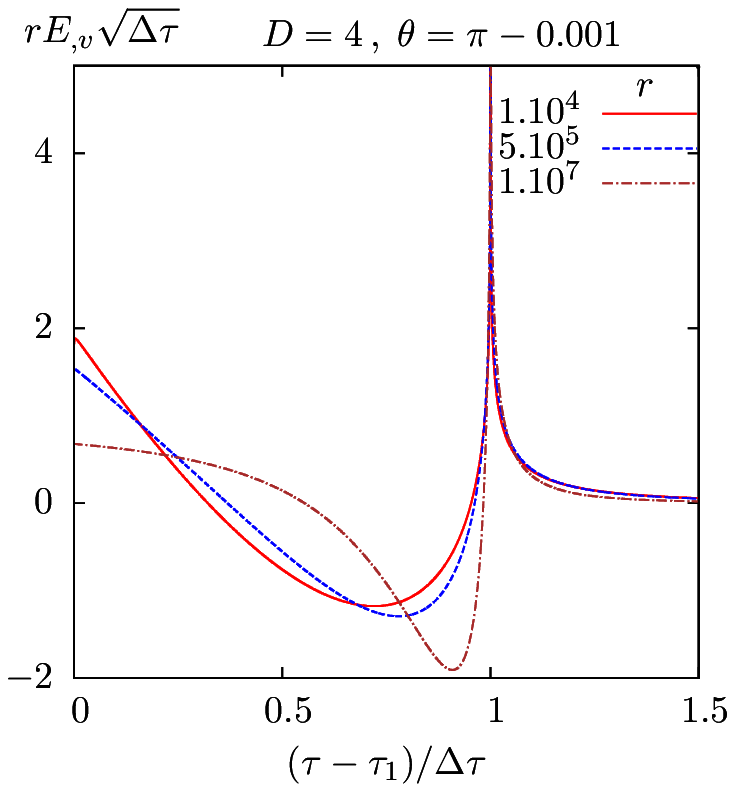} \hspace{1.5mm} 
\includegraphics[scale=0.66,clip=true,trim= 0 0 0 0]{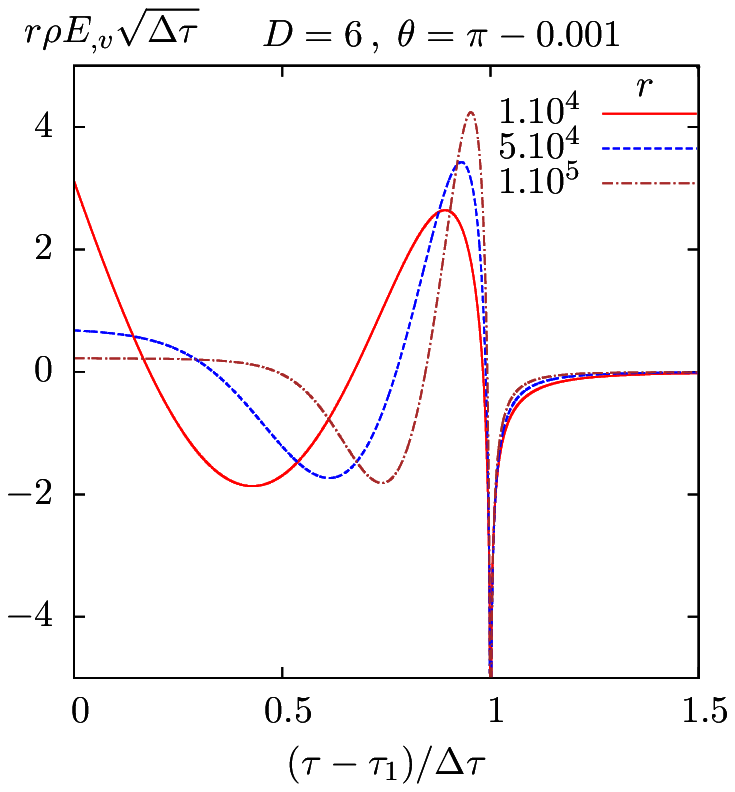} \hspace{1.5mm} 
\includegraphics[scale=0.66,clip=true,trim= 0 0 0 0]{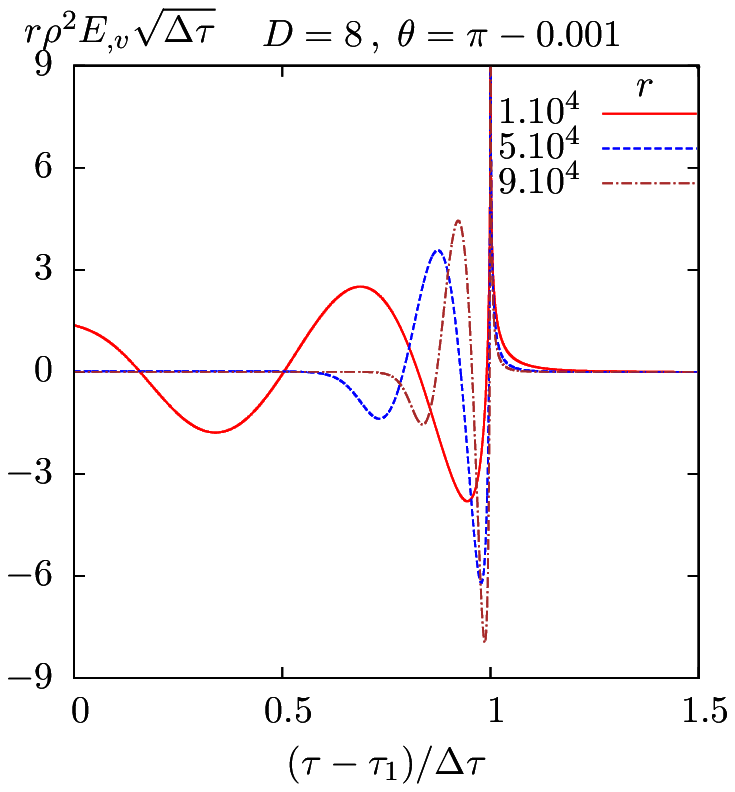}
\caption{\label{fig:RedthetaVarDvar} {\em Wave forms in a rescaled $\tau$ coordinate:} For all wave forms, the signal jumps to a non-zero value at $\tau=\tau_1$ and peaks at $\tau=\tau_2$. These are the values of $\tau$ at which rays 1 and 2 of Fig. \ref{rays} reach the observation point, respectively.}
\end{figure}

In Fig.~\ref{fig:rVarDvar}, we present some wave forms ($D$~even) as a function of $\tau$. 
Since we are interested in the limit near the negative $z$ axis (where our approximations are justified), we use examples with $\theta$ away from the axis by $0.01$ radians. A first observation is that we have shifted $\tau$ by $\tau_1$ for all wave forms such that zero corresponds to the beginning of the radiation burst. A consistency check is that, indeed, the numerically determined integration domain agrees exactly (within very small numerical errors) with this expectation. Thus all the wave forms have a sudden step at zero which arises naturally from the numerical code. Another feature which is verified numerically, is that the (integrable) singular peak of radiation observed in all plots, occurs exactly at $\tau=\tau_2$. This is explicitly exhibited in Fig. \ref{fig:RedthetaVarDvar}, where besides the shift of $\tau$ by $\tau_1$ we have rescaled the axis by $\Delta \tau=\tau_2-\tau_1$.

An advantage of our numerical method compared to~\cite{D'Eath:1992hb} is that no approximations were made in the integral. In fact in~\cite{D'Eath:1992hb}, the limit $r\rightarrow +\infty$ was taken inside the integral before actually performing the integration. This removes the $\rho'^2$ term in~\eqref{eq:domain}, which is responsible for excluding regions of the domain of integration with large $\rho'$. In four dimensions, due to the logarithmically growing $\Phi$ profile, it turns out that such regions of large $\rho'$ are indeed cut off from the integral by the $\ln \rho'$ term. Thus in the large $r$ limit, the wave form does approach the one provided in Fig.~4 of~\cite{D'Eath:1992hb}. This can be seen in the first plot as $r$ becomes large. However the approximation in~\cite{D'Eath:1992hb} makes it less clear when the radiation burst starts and peaks. In our approach we simply use the full numerical integral without further approximations, for finite $r$, and find the limit numerically. Finally, for $D>4$, the  procedure in~\cite{D'Eath:1992hb} is not valid, because it would amount to neglecting the only term growing with $\rho'$ in the definition of $x_\star$, so there would be no regularisation of the domain and the integral would not converge. For $D>4$, there is a notable difference as we take the large $r$ limit, namely that the radiation pulse becomes more concentrated around $\tau=\tau_2$. In fact, the same happens in four dimensions, however more slowly (as $r$ increases) due to the logarithmic nature of the profile of the shock wave~\footnote{This is clear by looking at the ranges of $r$ for the different curves in each plot where, in $D=4$, $r$ varies by several orders of magnitude, whereas for $D>4$ it only varies roughly by one order of magnitude and the separation between curves is similar.}. The other difference is that the wave form  acquires one more oscillation in each jump to the next $D$ even. Finally we note (without plotting explicitly) that the effect of decreasing $\theta$ (with $r$ fixed), is the same as when we decrease $r$ with $\theta$ fixed\footnote{This should not be surprising since the $r$ and $\theta$ dependence appears mostly in $\rho=r\sin\theta$ factors.}, that is, the radiation burst becomes sharper and occurs in a smaller period $\sim \Delta \tau$ as we move closer to the axis.

\begin{figure}[t]
\hspace{-6mm}\includegraphics[scale=0.62,clip=true,trim= 0 0 0 0]{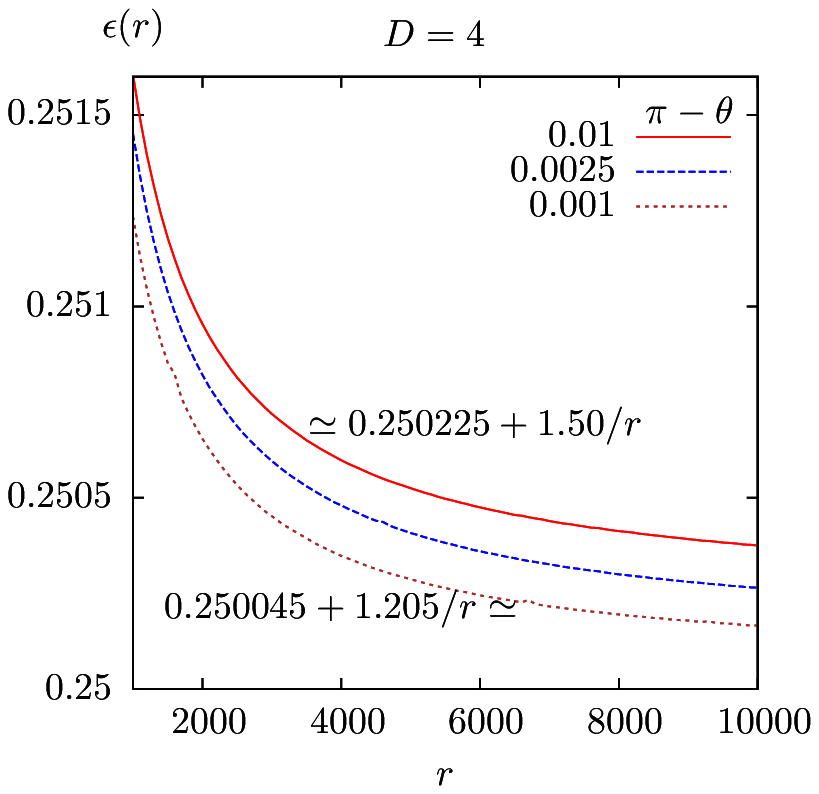} \hspace{2mm}
\includegraphics[scale=0.62,clip=true,trim= 0 0 0 0]{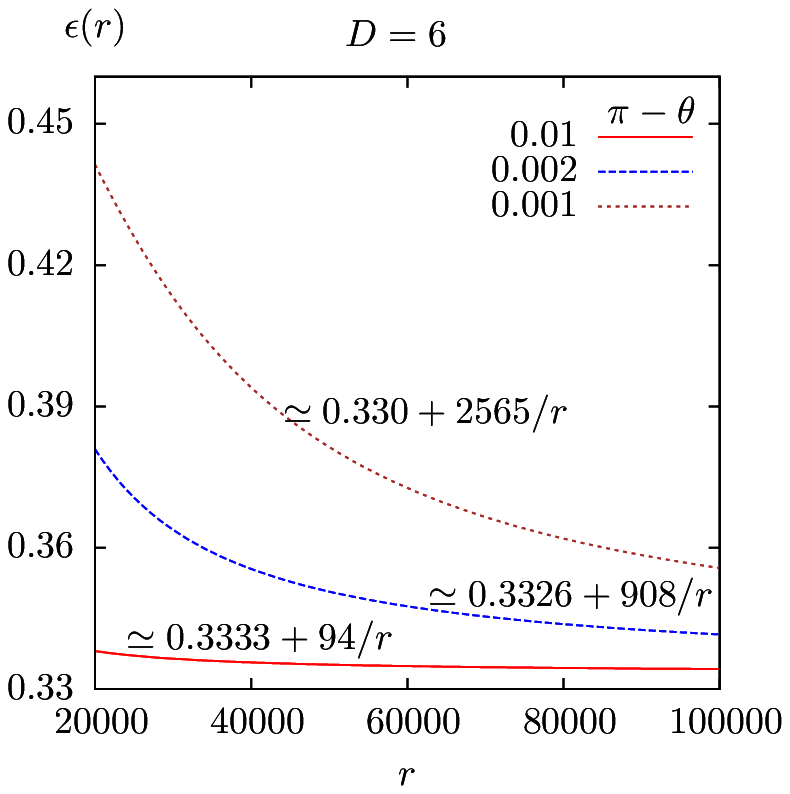} \hspace{0mm} 
\includegraphics[scale=0.62,clip=true,trim= 0 0 0 0]{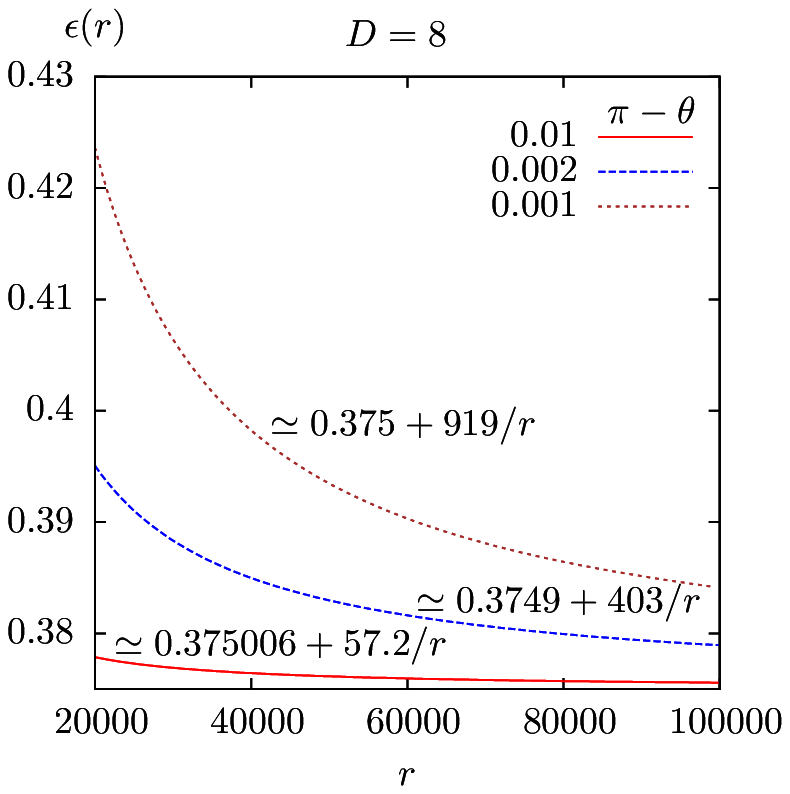} 
\caption{\label{fig:IntthetaVarDvar} {\em Limiting fractions:} These plots show $\epsilon(r)$, whose limit for $r$ large and small $\pi-\theta$ gives the estimate $\epsilon_{\rm radiated}$. The best estimate for the limit can be read from the constant term in the asymptotic form used to fit the numerical points of the red curves (for  which the angle is the smallest).}
\end{figure}

After we obtain the wave forms for a certain $r,\theta$, we can compute the estimate for the fraction of radiated energy, by computing the integral in~\eqref{rad1} and taking the limit $r\rightarrow +\infty$, $\hat\theta\rightarrow 0$. In Fig.~\ref{fig:IntthetaVarDvar}  
we show plots of the radiated fraction of energy as a function of $r$ for several $\theta$. We have performed a fit of each numerical curve to extract the limit, by using an asymptotic expansion of the form $\epsilon(r)\simeq \epsilon_{\rm radiated}+a/r$ (indicated next to each curve). Since we have computed this function for very large values of $r$ we obtain a fit which is basically indistinguishable from the numerical tails, so $\epsilon_{\rm radiated}$ is extracted with a relative error of less than $0.001$. Note that the difference in the shape of the curves for different $\theta$, is again smaller for the $D=4$ case due to the logarithmic dependence of the shock wave profile. Since we have used very small angles, all the asymptotic values in the different curves are consistent with the limiting values quoted in the introduction (within the relative precision of $0.001$ and the error from the $a/r$ correction).

\subsection{Discussion of the method}
The computation performed in this paper relies on perturbation theory. Why should perturbation theory be valid in describing a non-linear effect (like black hole formation) and an associated strong field effect, such as the generation of a strong burst of gravitational radiation? In \cite{D'Eath:1992hb} this is justified as follows. Mathematically, the boost provides a small parameter for expanding the geometry (the ratio of energies of weak to strong shock waves). Analysing~\eqref{fo}, one concludes that in the boosted frame the initial conditions are indeed perturbative except in a small vicinity of $\rho=0$ \text{and} sufficiently large $v>0$. This has a physical picture. The null generators of the weak shock wave (in the boosted frame) are not only bent towards the axis of the collision, as they cross the strong shock, but they also suffer a redshift which becomes larger as the null generator approaches the axis and infinitely large for generators at the axis. This is manifest in Fig. \ref{focusing4} and \ref{focusing}. The radiation that hits a spacetime point (not at the axis), therefore, comes first from the far field region of the collision (ray 1 in Fig. \ref{rays}), which is given a large head start as compared to its near field counterpart (ray 2 in Fig. \ref{rays}). Since gravity is weak in the far field region, perturbation theory should therefore be accurate in describing this part of the signal. Moreover, due to focusing, the overall amount of radiation measured near the axis, at very large distances, coming from the far field region need not be small. Therefore, it is concluded in \cite{D'Eath:1992hb} that this problem provides one example where perturbation theory can tackle a strong field effect.

The interpretation of the wave forms we have provided in this paper shows, however, that the above argument must be considered with care. The wave forms exhibited in Fig. \ref{fig:rVarDvar} and \ref{fig:RedthetaVarDvar} are dominated by the signal \textit{near} $\tau_2$, which is the signal that crosses the near field region. This indicates therefore, that there is a large contribution to the signal from outside the region where linear theory is clearly applicable and  a  correction from higher order terms in perturbation theory is expected. But the near field region experienced by ray 2 becomes smaller at very large $r$ and small $\hat{\theta}$. Furthermore, as we have learned in $D=4$, the first order perturbation provides a better estimate than the trapped surface argument and, more importantly, the second order perturbation provides a result consistent with the numerical simulations. We find this an important motivation to pursue this computation. It will be interesting to see if the agreement between numerical results and perturbation theory still holds in higher dimensions.

In order to compute the total radiation emitted, we have determined the radiation in the vicinity of $\theta=\pi$ and extrapolated it under an isotropy assumption for $dE/d\cos\theta$, following \cite{D'Eath:1992hb}. This is partly justified by the Zero Frequency Limit approximation \cite{Smarr:1977fy}, which predicts an isotropic angular distribution of gravitational radiation (with two blind spots along the symmetry axis) for the collision of two equal mass black holes as the speed of light is approached. Nevertheless, the second order computation of D'Eath and Payne introduces an angular dependence \cite{D'Eath:1992qu}, which seems to be crucial for a more accurate estimate.

One final remark concerning odd $D$. It is well known that the Green's functions of the D'Alembertian operator have support \textit{on} the light cone in even $D$ and both \textit{on} and \textit{inside} the light cone for odd $D$, cf. \eqref{greeneven} and \eqref{greenodd} (see, e.g. \cite{Galtsov:2001iv,Cardoso:2002pa,Barvinsky:2003jf}). This leads to the unfamiliar property that odd $D$ dimensional Minkowski space behaves as a dispersive medium. It may seem that the integral solution \eqref{eq:integral_sol} is missing this property for odd $D$, since it has support solely at the intersection of the past light cone with the initial data surface (as illustrated in  Fig. \ref{pastlightcone}). But it is not so. The equivalence between the integral solution and the Green's function method is shown in appendix \ref{apa}.

\section{Conclusions}
\label{section5}
In this paper we have provided an estimate for the energy radiated into gravitational waves in a collision of shock waves in a $D$ dimensional spacetime. These shock waves provide an intuitive description, due to the infinite Lorentz boost, of the gravitational field of ultra-relativistic particles. Moreover, they do capture their gravitational interactions. Indeed, the scattering amplitude for two (scalar) particles in the eikonal regime of perturbative quantum gravity \cite{Kabat:1992tb}, can be equivalently computed by analysing the scattering of a (test) plane wave in a shock wave background \cite{'tHooft:1987rb} (see also \cite{Lodone:2009qe}).  Thus our computation applies to particle collisions in a regime wherein their interaction is dominated by gravity, and well described by classical general relativity. This is the transplanckian scattering regime \cite{'tHooft:1987rb}.  

If the fundamental Planck scale is of the order of the TeV scale, a key issue for phenomenology is how good the classical and semi-classical approximations are to describe the black holes that could be produced in particle collisions. In other words how much of the initial centre of mass energy of the process stays in the final black hole. The answer essentially amounts to understand how much energy is lost in gravitational radiation. The best estimates for this radiated energy, to date, come from apparent horizon computations \cite{Eardley:2002re}. In four dimensions the apparent horizon estimates are off by a factor of two as compared to improved values obtained either from perturbative computations in colliding shock waves backgrounds \cite{D'Eath:1992hb,D'Eath:1992hd,D'Eath:1992qu} or numerical relativity results \cite{Sperhake:2008ga}. The final black hole is therefore more massive (and hence more classical) than what has been anticipated by apparent horizon estimates. A similar conclusion in higher dimensions would have phenomenological relevance. 

In this paper we have computed the metric in the future of the collision of two shock waves in first order perturbation theory and extracted the corresponding gravitational radiation. Although our result should only be faced as an estimate, what we know from $D=4$ indicates that it is an \textit{improved} estimate as compared to that obtained from the analysis of trapped surfaces \cite{Eardley:2002re}. 

The apparent horizon estimate gives the same trend for the $D$ dependence as our result. This trend is also observed in numerical relativity results for a low energy collision of black holes in $D=4$ and $D=5$: these show that the energy radiated increases, from $D=4$ to $D=5$ \cite{Witek:2010xi}. Curiously, another bound that applies to these low energy collisions exhibits an opposite trend. From the area law, that bound may be obtained, as first shown by Hawking \cite{Hawking:1971tu}, for the energy emitted into gravitational radiation in a collision of non-spinning, equal mass black holes starting from rest at infinite distance. This bound starts at precisely the same value, $29.3\%$, as the one discussed herein for high energy collisions, but it \textit{decreases} with dimension. Another context where a similar decrease with $D$ occurs, though in a different regime - that of extreme mass ratio, was studied in~\cite{Berti:2010gx,Berti:2003si,Cardoso:2005jq}, where the gravitational radiation emitted by point particles falling into a black hole was computed.

This paper also clarifies two aspects of the first order computation by D'Eath and Payne \cite{D'Eath:1992hb}. The first is the approximation procedure in determining the emitted power, namely neglecting $\rho'^2$ terms in the computation of \eqref{rad2}. Since we have evaluated \eqref{rad2} numerically, no such approximation was used, and our result is consistent with that of D'Eath and Payne within less than one percent. This shows that the approximation used in \cite{D'Eath:1992hb} is robust in $D=4$. The second is the time integration domain. It seems strange, at first sight, that there is a non-vanishing gravitational signal for all $\tau\in \mathbb{R}$ in $D=4$, in the limit $r\rightarrow \infty$ for the observation point. This is a consequence of the nature of the coordinate transformation from Brinkmann to Rosen in $D=4$ and the jump of the null generators of the weak shock to negative values of $v$ as they cross the strong shock (cf. Fig. \ref{focusing4}), for large $\rho$. In $D>4$ the coordinate transformation is different and so is the time integration, which has support only for $\tau\in \mathbb{R}^+$.  Moreover, the observation we have made concerning the matching of the peaks in the obtained wave forms with a simple ray analysis, is useful in understanding the shape of the wave forms, which was not discussed in  \cite{D'Eath:1992hb}.

Finally, we hope that this paper will help setting the stage for further use of this technique, since it can be applied to various generalisations of the problem we are considering. In particular, one immediate goal is to consider the second order in perturbation theory. We observe that the initial conditions are exact to second order which might be at the basis of the good agreement between the perturbative computation and the numerical relativity result in $D=4$.

\section*{Acknowledgements}
We would like to thank V. Cardoso, L. Gualtieri and U. Sperhake for helpful discussions. We are especially grateful to the referee of this paper for finding an incorrect $D$ dependent factor. C.H. would like to thank CERN, theory division, for hospitality, where part of this work was done. M.S. is supported by the FCT grant SFRH/BPD/69971/2010. This work is also supported by the grants CERN/FP/116341/2010, PTDC/FIS/098962/2008 and PTDC/FIS/098025/2008.

\appendix

\section{General integral solution}
\label{apa}
In this section we  generalise the integral solution found in \cite{D'Eath:1992hb} to higher dimensions. The most standard method is to use Green's functions. In our case, however, there is a simple way of finding the solution by using an ansatz inspired by the four dimensional case. First we provide such a derivation and  after we check it using the well know Green's functions method in $D$-dimensional Minkowski spacetime.

\subsection{Integral operator ansatz}
 Before attempting to generalise the integral solution of the wave equation $\square F=0$ in $u\geq 0$ with initial data at $u=0$ provided in \cite{D'Eath:1992hb}, let us start by investigating how it works in four dimensions. In that case the solution is
\begin{equation}
F(u,v,x_i)=\dfrac{1}{2\pi u}\int d^{2}x' \dfrac{\partial}{\partial v'}F(0,v',x_i') \label{eq:intSol4D} \ , 
\end{equation}
where, for each $x'$, $v'$ defines points, at $u=0$, \textit{on} the past light cone of the event $(u,v,x_i)$:
\begin{equation}
v'=v-\dfrac{|x-x'|^2}{2u} \; .
\end{equation}
We can compute partial derivatives of $F$ by acting directly on the explicit $u$ dependence, and the implicit dependence on $u,v,x_i$ through $v'$. Then using 
\begin{equation}
\dfrac{\partial}{\partial v}=\dfrac{\partial}{\partial v'} \;,  \hspace{1cm} \dfrac{\partial }{\partial u}=\dfrac{\partial v'}{\partial u}\dfrac{\partial}{\partial v'}=\dfrac{|x-x'|^2}{2 u^2}\dfrac{\partial}{\partial v'}\;,  \hspace{1cm}  \dfrac{\partial}{\partial x_i}=\dfrac{\partial v'}{\partial x_i}\dfrac{\partial}{\partial v'}=-\dfrac{x^i-x'^i}{u}\dfrac{\partial}{\partial v'} \ , 
\end{equation}
we can compute $\square F$ and check it is indeed zero. Furthermore, we need to check the initial condition. If we try to take the limit $u\rightarrow 0$ directly in~\eqref{eq:intSol4D} we realise that there is no easy way to deal with the $v'$ derivative. To remove it we can work in Fourier space by taking the Fourier transform with respect to $v$ (we denote it $\tilde F$ and replace the corresponding argument by $q$)
\begin{equation}
\tilde F(u,q,x_i)=\int d^{2}x' \dfrac{e^{-iq\frac{|x_i-x'_i|^2}{-2u}}iq}{-2\pi u} \tilde{F}(0,q,x_i') =\int d^{2}x' \delta_{\epsilon}(x_i-x'_i) \tilde{F}(0,q,x_i') \ , 
\end{equation}
where we have identified the parameter $\epsilon=-iu/q$ and a function $\delta_{\epsilon}$ which becomes the $2$-dimensional Dirac delta distribution in the $\epsilon\rightarrow 0$ limit. So, in Fourier space, the initial condition is obeyed and the same holds after Fourier inversion.

In higher dimensions we try a similar form
\begin{equation}
F(u,v,x_i)=\dfrac{1}{N u^{\alpha}}\int d^{D-2}x' O_{v'}F(0,v',x_i') \ , \label{eq:intSol}
\end{equation}
where for the moment we leave the overall normalisation $N$ arbitrary, $v'$ is as before, and $\alpha$ is to be determined as well as the $O_{v'}$ operator acting on $F$. We can again compute $\square F=0$ to obtain
\begin{equation}
(D-2-2\alpha)\dfrac{1}{N u^{\alpha+1}}\int d^{D-2}x' \partial_{v'}O_{v'}F(0,v',x_i')=0 \; .
\end{equation}
Thus this integral ansatz must have $\alpha=(D-2)/2$. Incidentally, this is exactly the correct number of $u$ powers needed to obtain a $D-2$~dimensional delta function by repeating the reasoning for the initial condition. Again, we Fourier transform with respect to $v$ to obtain
\begin{eqnarray}
\tilde F(u,q,x_i)&=&\dfrac{(2\pi)^{\frac{D-2}{2}}}{N}\int d^{D-2}x'\dfrac{e^{-iq\frac{|x-x'|^2}{-2u}}}{ (2\pi u)^{\frac{D-2}{2}}}\int dv' e^{iqv'} O_{v'}F(0,v',x_i') \nonumber\\
&=&\dfrac{(2\pi)^{\frac{D-2}{2}}}{N}\int d^{D-2}x'\delta_\epsilon(x-x')(-iq)^{-\frac{D-2}{2}}\int dv' e^{iqv'} O_{v'}F(0,v',x_i') \; . \;
\end{eqnarray}
So in the $u\rightarrow 0$ limit we obtain
\begin{eqnarray}
\tilde F(0,q,x_i)=\dfrac{(2\pi)^{\frac{D-2}{2}}}{N}\dfrac{1}{(-iq)^{\frac{D-2}{2}}}\int dv' e^{iqv'} O_{v'}F(0,v',x_i) \ . \label{initcond}
\end{eqnarray}
In higher dimensions, for the right hand side to reduce to the left hand side, we need a generalisation of the Fourier transform of the operator $O_{v'}\rightarrow \partial^{(D-2)/2}_{v'} $. This is defined through its Fourier transform when acting on a function $f(v)$ (so it is a distributional operator)
\begin{equation}
\int e^{iqv}O_v f(v)=(-iq)^{\frac{D-2}{2}}\tilde f(q) \; .
\label{Ofourier}
\end{equation} 
Using~\eqref{Ofourier} and choosing $N=(2\pi)^{(D-2)/2}$, then~\eqref{initcond} is solved. In even dimensions $O_v$ is just a partial derivative. However in odd dimensions we have a fractional partial derivative, thus formally, the general integral solution is
\begin{equation}
\label{sol:intEqapp}
F(u,v,x_i)=\dfrac{1}{(2\pi u)^{\frac{D-2}{2}}}\int d^{D-2}x'\partial_{v'}^{\frac{D-2}{2}} F(0,v',x_i')  \; .
\end{equation}

\subsection{Green's function solution}
A more standard derivation, consists of reducing our Cauchy problem with initial conditions on the null hypersurface $u=0$, to a wave equation problem for a distribution with a source term. Then the general solution in $D$~dimensions can be found, for example, from Theorem~6.3.1 of~\cite{Friedlander:112411}
\begin{equation}
F(u,v,x_i)=2\int du'dv'd^{D-2}x' \, G(u-u',v-v',x_i-x_i')\delta(u)\partial_{v'}F(0,v',x_i') \; .
\end{equation}
The Green's function for even $D$ is
\begin{equation}
\label{greeneven}
G(u,v,x_i)=\dfrac{1}{2(2\pi u)^{\frac{D-2}{2}}}\delta^{(\frac{D-2}{2})}\left(\frac{|x|^2}{2u}-v\right) \ ,
\end{equation}
where the superscript denotes denotes the distributional derivative of order $(D-2)/2$, of the Dirac delta distribution. For odd $D$
\begin{equation}
G(u,v,x_i)=\dfrac{1}{2(2\pi u)^{\frac{D-2}{2}}\Gamma\left(\frac{4-D}{2}\right)}\dfrac{\theta^{(\frac{D-2}{2})}\left(\frac{|x|^2}{2u}-v\right)}{\left(\frac{|x|^2}{2u}-v\right)^{\frac{D-2}{2}}} \ , \label{greenodd}
\end{equation}
where this distribution is to be understood as giving the finite part of the integral when acting on functions. In even dimensions, we can perform the $u',v'$ integration immediately and use the action of the distributional derivative of the delta function to get exactly~\eqref{sol:intEqapp}. For odd $D$, we  first note that by working in Fourier space, we can define the negative order of the delta distribution, ($p<0$) as
\begin{equation}
\delta^{(p)}(x)=\dfrac{1}{\Gamma(-p)}x^{-p-1}\theta(x) \; .
\end{equation}
This definition can be used to extend (recursively) to positive orders, by acting with derivatives on $\delta^{(p)}(x)$. For example, the fractional derivative of the delta function of order $1/2$, is simply
\begin{eqnarray}\label{eq:delta12}
\delta^{(1/2)}(x)&=&\dfrac{d}{dx}\delta^{(-1/2)}(x)=\dfrac{d}{dx}\left[\dfrac{1}{\Gamma(\frac{1}{2})}|x|^{-1/2}\theta(x)\right]=-\dfrac{1}{\sqrt{\pi|x|}}\left[\dfrac{\theta(x)}{2x}-\delta(x)\right]  \ .
\end{eqnarray}
and any higher order fractional derivative can be defined similarly. We note that up to singular delta function terms which we must discard (due to the finite part prescription for its integral with a function), we can write the $D$ odd Green's function exactly as~\eqref{greeneven} with the fractional order delta distribution. Thus the same integral solution, with a fractional partial derivative, holds in odd dimensions in agreement with the first derivation.

\section{Solution for the transverse components}
\label{apb}

In Fourier space (with respect to $v$), the solution of~\eqref{hareq2} for the spatial components of the metric perturbations reads
\begin{equation}
\tilde h^{N}_{ij}(u,q,x_k)=\dfrac{1}{(2\pi u)^{\frac{D-2}{2}}}\int d^{D-2}x' e^{iq\frac{|x-x'|^2}{2u}}(-iq)^{\frac{D-2}{2}}\tilde h^{N}_{ij}(0,q,x_k')  \; .
\end{equation}
For convenience we write the initial perturbation as to separate the $v$ dependent term:
\begin{equation}
h^{N}_{ij}(0,v,x_k)=\left(\delta_{ij}-(D-2)\frac{x_ix_j}{\rho^2}\right)\dfrac{\Phi'}{\rho}\left(\sqrt{2}v-\Phi\right)\theta\left(\sqrt{2}v-\Phi \right)  \ .
\end{equation}
Then the Fourier transform of the initial perturbation is obtained as 
\begin{equation}
\tilde h^{N}_{ij}(u,q,x_k)=\dfrac{\sqrt{2}(-iq)^{\frac{D-6}{2}}}{(2\pi u)^{\frac{D-2}{2}}}\int d^{D-2}x' \left(\delta_{ij}-(D-2)\frac{x'_ix'_j}{\rho'^2}\right)\dfrac{\Phi'(\rho')}{\rho'}e^{iq\left(\frac{|x-x'|^2}{2u}+\frac{\Phi(\rho')}{\sqrt{2}}\right)} \; .
\end{equation}
The integration measure and $\rho'$ are invariant under rotations. So we define a rotation, such that $y_i=R_{ij}x_j$, is aligned with the $y_1$ direction and has magnitude $\rho$. Similarly $y'_i=R_{ij}x'_j$, so then
\begin{multline}
\tilde h^{N}_{ij}(u,q,x_k)= \\ R^{-1}_{i\ell}R^{-1}_{jm}\dfrac{\sqrt{2}(-iq)^{\frac{D-6}{2}}}{(2\pi u)^{\frac{D-2}{2}}}\int d^{D-2}y' \left(\delta_{\ell m}-(D-2)\frac{y'_\ell y'_m}{\rho'^2}\right)\dfrac{\Phi'(\rho')}{\rho'}e^{iq\left(\frac{\rho^2-2\rho\rho'\cos\theta'+\rho'^2}{2u}+\frac{\Phi(\rho')}{\sqrt{2}}\right)} \; ,
\end{multline}
where $\theta'$ projects the vector $y'$ onto the $y'_1$ direction. In $D=4$, $\theta'\in [0,2\pi]$ whereas for $D>4$ $\theta'\in [0,\pi]$.
The parity of the integrand with respect to $y'_2,\ldots,y'_{D-2}$ is always well defined (there is only no definite parity with respect to $y'_1$ due to the exponential). Thus only the $\ell=m$ components are non-zero, either due to the $\delta_{\ell m}$ or because if $\ell \neq m$, then $y'_\ell y'_m$ is odd when integrated over $y'_2,\ldots,y'_{D-2}$. Changing to hyperspherical coordinates 
\begin{eqnarray}
&&\tilde h^{N}_{ij}(u,q,x_k)= R^{-1}_{i\ell}R^{-1}_{jm}\dfrac{\sqrt{2}(-iq)^{\frac{D-6}{2}}}{(2\pi u)^{\frac{D-2}{2}}}  \\  &&\int d\rho' \, \Phi'(\rho')\rho'^{D-4} \hspace{-1mm}\int\hspace{-1mm} d\theta' (\sin\theta')^{D-4} e^{iq\left(\frac{\rho^2-2\rho\rho'\cos\theta'+\rho'^2}{2u}+\frac{\Phi(\rho')}{\sqrt{2}}\right)} \hspace{-1mm} \int \hspace{-1mm} d\Omega_{D-4}\left[\delta_{\ell m}-(D-2)\frac{y'_\ell y'_m}{\rho'^2}\right] . \nonumber
\end{eqnarray}
The angular integrals are obtained straightforwardly using standard integrals on the $(D-2)-$sphere, so we obtain
\begin{multline}
\tilde h^{N}_{ij}(u,q,x_k)= R^{-1}_{i\ell}R^{-1}_{jm}\dfrac{\sqrt{2}(-iq)^{\frac{D-6}{2}}}{(2\pi u)^{\frac{D-2}{2}}} \\ \int d\rho' \, \Phi'(\rho')\rho'^{D-4}\int_0^\pi d\theta' (\sin\theta')^{D-4} e^{iq\left(\frac{\rho^2-2\rho\rho'\cos\theta'+\rho'^2}{2u}+\frac{\Phi(\rho')}{\sqrt{2}}\right)} \Omega_{D-4}G_{\ell m}(\cos\theta') \;  ,
\end{multline}
with
\begin{equation}\label{eq.Gij}
G_{\ell m}(\cos\theta')=\delta_{\ell m} -{\rm diag}\left\{(D-2)\cos^2\theta',\frac{D-2}{D-3}\sin^2\theta',\ldots,\frac{D-2}{D-3}\sin^2\theta'\right\} \; ,
\end{equation}
where we have defined $\Omega_0=2$ to account for the fact that $\theta'$ is a polar angle in $D=4$. Note that the first  angular integral (corresponding to $y_1'y_1'$) is just an area factor and the other components can be obtained using the vanishing of the trace. This last property is satisfied by the final result~\eqref{eq.Gij} which works as an independent check of the angular integrals. Fourier inversion then yields
\begin{eqnarray}
 &&h^{N}_{ij}(u,v,x_k)= R^{-1}_{i\ell}R^{-1}_{jm}\dfrac{\sqrt{2}\Omega_{D-4}}{(2\pi u)^{\frac{D-2}{2}}} \\ &&\int \hspace{-1mm} d\rho' \, \Phi'(\rho')\rho'^{D-4}\hspace{-1mm}\int_{0}^{\pi}\hspace{-1mm} d\theta' (\sin\theta')^{D-4}G_{\ell m}(\cos\theta')\delta^{\left(\frac{D-6}{2}\right)}\hspace{-1mm}\left(v-\frac{\rho^2-2\rho\rho'\cos\theta'+\rho'^2}{2u}-\frac{\Phi(\rho')}{\sqrt{2}}\right) . \nonumber
\end{eqnarray}
 The exponent of the delta denotes the order of the derivative, which can be fractional or negative as in~\eqref{eq:delta12}.
Note that due to the vanishing of the trace, actually only one component of the $ij$ is independent. If we factor out the rotations, we can define the rotated $11$ perturbation as 
\begin{multline}
 h^{rot}_{11}(u,v,\rho)\equiv E(u,v,\rho)=-\dfrac{\sqrt{2}\Omega_{D-4}}{(2\pi u)^{\frac{D-2}{2}}}\int d\rho'\, \rho'^{D-4} \Phi'(\rho') \times \\ \times \int_0^\pi d\theta' (\sin\theta')^{D-4}\left((D-2)\cos^2\theta'-1\right) \delta^{(\frac{D-6}{2})}\left(v-\frac{\rho^2-2\rho\rho'\cos\theta'+\rho'^2}{2u}-\frac{\Phi(\rho')}{\sqrt{2}}\right)
 \; . 
\end{multline}
Now if we define $x\equiv \cos\theta'$, take a $v$ derivative and use the scaling properties of the derivative of the delta distribution 
\begin{equation}
 E_{,v}=\dfrac{\sqrt{8}\Omega_{D-4}}{(2\pi \rho)^{\frac{D-2}{2}}(D-1)} \int_{0}^{+\infty} \dfrac{d\rho'}{\rho'^{\frac{D}{2}}}\,\int_{-1}^{1} dx \,\dfrac{d^2}{dx^2}\left[(1-x^2)^{\frac{D-1}{2}}\right] \delta^{(\frac{D-4}{2})}\left(x-x_\star\right) 
 \; ,
\end{equation}
with
\begin{equation}
x_\star \equiv \dfrac{U\Phi\left(\rho'\right)+\rho'^2-UT}{2\rho \rho'} \; ,
\end{equation}
where $U\equiv \sqrt{2}u$ and $T\equiv \sqrt{2}v-\rho^2/U$.
To perform the integral, we can always integrate by parts $M=[(D-4)/2]$ times (the brackets indicate the integer part) without obtaining any boundary terms\footnote{This is because terms of the form $(1-x^2)^{p}$ are  always present (evaluating to zero at the boundaries) and  the delta functions only have support on the boundary for a subset of points with measure zero.}, so we end up with 
\begin{equation}
 E_{,v}=\dfrac{\sqrt{8}\Omega_{D-4}(-1)^{M}}{(2\pi \rho)^{\frac{D-2}{2}}(D-1)} \int_{0}^{+\infty} \dfrac{d\rho'}{\rho'^{\frac{D}{2}}}\,\int_{-1}^{1} dx \,\dfrac{d^{M+2}}{dx^{M+2}}\left[(1-x^2)^{\frac{D-1}{2}}\right] \delta^{(\frac{q}{2})}\left(x-x_\star\right) 
 \; ,
\end{equation}
where $q=0$ for $D$ even and $q=1$ for $D$ odd. Then in even dimensions (note $\mathcal{D}$ is defined such that $-1\le x_\star \le 1$)
\begin{equation}
 E_{,v}=\dfrac{\sqrt{8}\Omega_{D-4}(-1)^{M}}{(2\pi \rho)^{M+1}(D-1)} \int_{\mathcal{D}} \dfrac{d\rho'}{\rho'^{M+2}} \,\dfrac{d^{M+2}}{dx^{M+2}}\left[(1-x^2)^{M+\frac{3}{2}}\right]_{x=x_\star} 
 \; ,
\end{equation}
and in odd dimensions, using~\eqref{eq:delta12},
\begin{equation}
 E_{,v}=\dfrac{\sqrt{8}\Omega_{D-4}(-1)^{M+1}}{(2\pi \rho)^{M+\frac{3}{2}}\sqrt{\pi}(D-1)} \int_{0}^{+\infty} \dfrac{d\rho'}{\rho'^{M+\frac{5}{2}}} \int_{-1}^{1} dx \,\dfrac{d^{M+2}}{dx^{M+2}}\left[(1-x^2)^{M+2}\right] \dfrac{d}{dx}\left[\dfrac{\theta\left(x - x_\star \right)}{\sqrt{x-x_\star}}\right] 
 \, .
\end{equation} The $x$ integral can be performed by expanding the polynomials and integrating by parts. Then if we define the following polynomials
\begin{equation}
\dfrac{d^{M+2}}{dx^{M+2}}\left[(1-x^2)^{M+2}\right]\equiv \sum_{k=0}^{M+2}\dfrac{c_k}{k!} x^k \; ,
\end{equation}
\begin{equation}
Q^{(M+2)}(x)\equiv \sqrt{1-x^2} \dfrac{d^{M+2}}{dx^{M+2}}\left[(1-x^2)^{M+\frac{3}{2}}\right]= \sum_{k=0}^{M+2}\dfrac{d_k}{k!} x^k \; ,
\end{equation}
\begin{equation}
P^{(M+2)}(x)\equiv \sum_{k=0}^{M+2}\sum_{j=0}^k\dfrac{ c_k x_\star^{k-j}(1-x_\star)^{j}}{(k-j)!j!(2j-1)} \; ,
\end{equation}
we obtain~\eqref{odd} in odd dimensions and~\eqref{even} in even dimensions.


\begin{thebibliography}{99}
\baselineskip=14pt


  
\bibitem{ArkaniHamed:1998rs}
  N.~Arkani-Hamed, S.~Dimopoulos, G.~R.~Dvali,
  ``The Hierarchy problem and new dimensions at a millimeter,''
  Phys.\ Lett.\  {\bf B429 } (1998)  263-272.
  [hep-ph/9803315].

\bibitem{Dimopoulos:2001hw}
  S.~Dimopoulos, G.~L.~Landsberg,
  ``Black holes at the LHC,''
  Phys.\ Rev.\ Lett.\  {\bf 87 } (2001)  161602.
  [hep-ph/0106295].
  
\bibitem{Giddings:2001bu}
  S.~B.~Giddings, S.~D.~Thomas,
  ``High-energy colliders as black hole factories: The End of short distance physics,''
  Phys.\ Rev.\  {\bf D65 } (2002)  056010.
  [hep-ph/0106219].

\bibitem{Khachatryan:2010wx}
  V.~Khachatryan {\it et al.} [ CMS Collaboration ],
  ``Search for Microscopic Black Hole Signatures at the Large Hadron Collider,''
  Phys.\ Lett.\  {\bf B697 } (2011)  434-453.
  [arXiv:1012.3375 [hep-ex]].

\bibitem{Aad:2009wy}
  G.~Aad {\it et al.}  [The ATLAS Collaboration],
  ``Expected Performance of the ATLAS Experiment - Detector, Trigger and Physics,'' [arXiv:0901.0512 [hep-ex]].
  

\bibitem{ATLAS-CONF-2011-065}
  The ATLAS Collaboration,
  ``Search for strong gravity effects in same-sign dimuon final states,''  CERN report ATLAS-CONF-2011-065, April 2011.
  
  \bibitem{ATLAS-CONF-2011-068}
  The ATLAS Collaboration,
  ``Search for Microscopic Black Holes in Multi-Jet Final
States with the ATLAS Detector at $\sqrt{s}$ = 7 TeV,''  CERN report ATLAS-CONF-2011-068, May 2011.
  
\bibitem{Park:2011je}
  S.~C.~Park,
  ``Critical comment on the recent microscopic black hole search at the LHC,''
    [arXiv:1104.5129 [hep-ph]].
  
\bibitem{Frost:2009cf}
  J.~A.~Frost, J.~R.~Gaunt, M.~O.~P.~Sampaio, M.~Casals, S.~R.~Dolan, M.~A.~Parker, B.~R.~Webber,
  ``Phenomenology of Production and Decay of Spinning Extra-Dimensional Black Holes at Hadron Colliders,''
  JHEP {\bf 0910 } (2009)  014.
  [arXiv:0904.0979 [hep-ph]].
  
\bibitem{Dai:2007ki}
  D.~-C.~Dai, G.~Starkman, D.~Stojkovic, C.~Issever, E.~Rizvi, J.~Tseng,
  ``BlackMax: A black-hole event generator with rotation, recoil, split branes, and brane tension,''
  Phys.\ Rev.\  {\bf D77 } (2008)  076007.
  [arXiv:0711.3012 [hep-ph]].
  
\bibitem{'tHooft:1987rb}
  G.~'t Hooft,
  ``Graviton Dominance in Ultrahigh-Energy Scattering,''
  Phys.\ Lett.\  {\bf B198 } (1987)  61-63.
  
\bibitem{D'Eath:1992hb}
  P.~D.~D'Eath, P.~N.~Payne,
  ``Gravitational radiation in high speed black hole collisions. 1. Perturbation treatment of the axisymmetric speed of light collision,''
  Phys.\ Rev.\  {\bf D46 } (1992)  658-674.

\bibitem{D'Eath:1992hd}
  P.~D.~D'Eath, P.~N.~Payne,
  ``Gravitational radiation in high speed black hole collisions. 2. Reduction to two independent variables and calculation of the second order news function,''
  Phys.\ Rev.\  {\bf D46 } (1992)  675-693.
  
\bibitem{D'Eath:1992qu}
  P.~D.~D'Eath, P.~N.~Payne,
  ``Gravitational radiation in high speed black hole collisions. 3. Results and conclusions,''
  Phys.\ Rev.\  {\bf D46 } (1992)  694-701.

\bibitem{Aichelburg:1970dh}
P.~C.~Aichelburg and R.~U.~Sexl,
``On The Gravitational Field Of A Massless Particle,''
Gen.\ Rel.\ Grav.\  {\bf 2}, 303 (1971).

\bibitem{Eardley:2002re}
  D.~M.~Eardley, S.~B.~Giddings,
  ``Classical black hole production in high-energy collisions,''
  Phys.\ Rev.\  {\bf D66 } (2002)  044011.
  [gr-qc/0201034].
  
\bibitem{Yoshino:2002tx}
  H.~Yoshino, Y.~Nambu,
  ``Black hole formation in the grazing collision of high-energy particles,''
  Phys.\ Rev.\  {\bf D67 } (2003)  024009.
  [gr-qc/0209003].
  
\bibitem{D'Eath:1976ri}
  P.~D.~D'Eath,
  ``High Speed Black Hole Encounters and Gravitational Radiation,''
  Phys.\ Rev.\  {\bf D18 } (1978)  990.
  
  
\bibitem{Rychkov:2004sf}
  V.~S.~Rychkov,
  ``Black hole production in particle collisions and higher curvature gravity,''
  Phys.\ Rev.\  {\bf D70 } (2004)  044003.
  [hep-ph/0401116].
  
\bibitem{Yoshino:2005hi}
  H.~Yoshino, V.~S.~Rychkov,
  ``Improved analysis of black hole formation in high-energy particle collisions,''
  Phys.\ Rev.\  {\bf D71 } (2005)  104028.
  [hep-th/0503171].

\bibitem{Sperhake:2008ga}
  U.~Sperhake, V.~Cardoso, F.~Pretorius, E.~Berti, J.~A.~Gonzalez,
  ``The High-energy collision of two black holes,''
  Phys.\ Rev.\ Lett.\  {\bf 101 } (2008)  161101.
  [arXiv:0806.1738 [gr-qc]].
  
\bibitem{Gal'tsov:2009zi}
  D.~V.~Gal'tsov, G.~Kofinas, P.~Spirin, T.~N.~Tomaras,
  ``Transplanckian bremsstrahlung and black hole production,''
  Phys.\ Lett.\  {\bf B683 } (2010)  331-334.
  [arXiv:0908.0675 [hep-ph]].

\bibitem{Constantinou:2011ju}
  Y.~Constantinou, D.~Gal'tsov, P.~Spirin, T.~N.~Tomaras,
  ``Scalar Bremsstrahlung in Gravity-Mediated Ultrarelativistic Collisions,''
  [arXiv:1106.3509 [hep-th]].
  
\bibitem{Sperhake:2009jz}
  U.~Sperhake, V.~Cardoso, F.~Pretorius, E.~Berti, T.~Hinderer, N.~Yunes,
  ``Cross section, final spin and zoom-whirl behavior in high-energy black hole collisions,''
  Phys.\ Rev.\ Lett.\  {\bf 103 } (2009)  131102.
  [arXiv:0907.1252 [gr-qc]].
  
\bibitem{Tangherlini:1963bw}
  F.~R.~Tangherlini,
  ``Schwarzschild field in n dimensions and the dimensionality of space problem,''
  Nuovo Cim.\  {\bf 27 } (1963)  636-651.
  
  \bibitem{Brinkmann}
  H.~W.~Brinkmann,
``Einstein spaces which are mapped conformally on each other,'' 
Math. \ Ann.\  {\bf 18} (1925) 119; also Proc. \ Natl. \ Acad. \ Sci. U.S. {\bf 9} (1923) 1.

\bibitem{Rosen}
A.~Einstein, N.~ Rosen,
``On gravitational waves'' J. Franklin Inst. {\bf 223} (1927) 43.

\bibitem{landau}
L.~D.~Landau, E.~M.~Lifshitz,
``The Classical Theory of Fields," Course of Theoretical Physics. Vol. 2, Butterworth Heinemann, 4th Ed. 1975, Reprinted 2000.

\bibitem{Zilhao:2010sr}
  M.~Zilhao, H.~Witek, U.~Sperhake, V.~Cardoso, L.~Gualtieri, C.~Herdeiro, A.~Nerozzi,
  ``Numerical relativity for D dimensional axially symmetric space-times: formalism and code tests,''
  Phys.\ Rev.\  {\bf D81 } (2010)  084052.
  [arXiv:1001.2302 [gr-qc]].
  
\bibitem{Witek:2010xi}
  H.~Witek, M.~Zilhao, L.~Gualtieri, V.~Cardoso, C.~Herdeiro, A.~Nerozzi, U.~Sperhake,
  ``Numerical relativity for D dimensional space-times: head-on collisions of black holes and gravitational wave extraction,''
  Phys.\ Rev.\  {\bf D82 } (2010)  104014.
  [arXiv:1006.3081 [gr-qc]].
  
\bibitem{Witek:2010az}
  H.~Witek, V.~Cardoso, L.~Gualtieri, C.~Herdeiro, U.~Sperhake, M.~Zilhao,
  ``Head-on collisions of unequal mass black holes in D=5 dimensions,''
  Phys.\ Rev.\  {\bf D83 } (2011)  044017.
  [arXiv:1011.0742 [gr-qc]].

\bibitem{Yoshino:2009xp}
  H.~Yoshino, M.~Shibata,
  ``Higher-dimensional numerical relativity: Formulation and code tests,''
  Phys.\ Rev.\  {\bf D80}, 084025 (2009).
  [arXiv:0907.2760 [gr-qc]].
  
  \bibitem{wald}
  R.~Wald,
  ``General Relativity,"
  The University of Chicago Press, 1984, Sec. 4.4.
  
\bibitem{Smarr:1977fy}
  L.~Smarr,
  ``Gravitational Radiation from Distant Encounters and from Headon Collisions of Black Holes: The Zero Frequency Limit,''
  Phys.\ Rev.\  {\bf D15 } (1977)  2069-2077.
  
\bibitem{Galtsov:2001iv}
  D.~V.~Galtsov,
  ``Radiation reaction in various dimensions,''
  Phys.\ Rev.\  {\bf D66 } (2002)  025016.
  [hep-th/0112110].
  
\bibitem{Cardoso:2002pa}
  V.~Cardoso, O.~J.~C.~Dias, J.~P.~S.~Lemos,
  ``Gravitational radiation in D-dimensional space-times,''
  Phys.\ Rev.\  {\bf D67 } (2003)  064026.
  [hep-th/0212168].
  
\bibitem{Barvinsky:2003jf}
  A.~O.~Barvinsky, S.~N.~Solodukhin,
  ``Echoing the extra dimension,''
  Nucl.\ Phys.\  {\bf B675 } (2003)  159-178.
  [hep-th/0307011].
  
\bibitem{Kabat:1992tb}
  D.~N.~Kabat, M.~Ortiz,
  ``Eikonal quantum gravity and Planckian scattering,''
  Nucl.\ Phys.\  {\bf B388 } (1992)  570-592.
  [hep-th/9203082].
  
\bibitem{Lodone:2009qe}
  P.~Lodone, V.~S.~Rychkov,
  ``Radiation Problem in Transplanckian Scattering,''
  JHEP {\bf 0912 } (2009)  036.
  [arXiv:0909.3519 [hep-ph]].
  
\bibitem{Hawking:1971tu}
  S.~W.~Hawking,
  ``Gravitational radiation from colliding black holes,''
  Phys.\ Rev.\ Lett.\  {\bf 26 } (1971)  1344-1346.

\bibitem{Berti:2010gx}
  E.~Berti, V.~Cardoso and B.~Kipapa, ``Up to eleven: radiation from particles with arbitrary energy falling into higher-dimensional black holes,''
  Phys.\ Rev.\  D {\bf 83} (2011) 084018
  [arXiv:1010.3874 [gr-qc]].

\bibitem{Berti:2003si}
  E.~Berti, M.~Cavaglia and L.~Gualtieri, ``Gravitational energy loss in high-energy particle collisions: Ultrarelativistic plunge into a multidimensional black hole,''
  Phys.\ Rev.\  D {\bf 69} (2004) 124011
  [arXiv:hep-th/0309203].

\bibitem{Cardoso:2005jq}
  V.~Cardoso, E.~Berti and M.~Cavaglia, ``What we (don't) know about black hole formation in high-energy collisions,''
  Class.\ Quant.\ Grav.\  {\bf 22} (2005) L61
  [arXiv:hep-ph/0505125].

\bibitem{Friedlander:112411}
  F.~G.~Friedlander,
  ``The wave equation on a curved space-time,''
 Cambridge\ Univ.\ Press (1975).


\end{thebibliography}
\end{document}